  \documentclass[referee]{aa} 
      \usepackage{graphicx}
      \usepackage[authoryear]{natbib}
      \bibpunct{(}{)}{;}{a}{}{,} 
      
\newcommand{\ea}{et al.}

\newcommand{\mags}{\>{\rm mag}}

\newcommand{\kms}{\>{\rm km}\,{\rm s}^{-1}}
\newcommand{\kkms}{\>{\rm K}\,{\rm km}\,{\rm s}^{-1}}
\newcommand{\kkk}{{\rm K}\,{\rm km}\,{\rm s}^{-1}}
\newcommand{\mpas}{\>{\rm mags\,arcsec}^{-2}}
\newcommand{\kk}{\>{\rm K}}
\newcommand{\pc}{\>{\rm pc}}
\newcommand{\kpc}{\>{\rm kpc}}
\newcommand{\mpc}{\>{\rm Mpc}}
\newcommand{\m}{\>{\rm m}}

\newcommand{\cm}{\>{\rm cm}}
\newcommand{\mm}{\>{\rm mm}}

\newcommand{\mhz}{\>{\rm MHz}}
\newcommand{\ghz}{\>{\rm GHz}}
\newcommand{\myr}{\>{\rm Myr}}
\newcommand{\gyr}{\>{\rm Gyr}}
\newcommand{\msun}{\>{\rm M_{\odot}}}

\newcommand{\as}{^{\prime\prime}}

\newcommand{\bdm}{\begin{displaymath}}
\newcommand{\edm}{\end{displaymath}}
\newcommand{\beq}{\begin{equation}}
\newcommand{\eeq}{\end{equation}}
\newcommand{\bit}{\begin{itemize}}
\newcommand{\eit}{\end{itemize}}
\newcommand{\ben}{\begin{enumerate}}
\newcommand{\een}{\end{enumerate}}
\newcommand{\bfi}{\begin{figure}[htb]}
\newcommand{\bpfi}{\begin{figure}[p]}

\newcommand{\htwo}{$\rm H_2$}
\newcommand{\coone}{$\rm ^{12}CO(1-0)$}
\newcommand{\cotwo}{$\rm ^{12}CO(2-1)$}
\newcommand{\ione}{$I_{10}$}
\newcommand{\itwo}{$I_{21}$}
\newcommand{\one}{$(1-0)$}
\newcommand{\two}{$(2-1)$}
\newcommand{\nc}{nuclear cluster}
\newcommand{\sbp}{surface brightness profile}
\newcommand{\mhi}{$M_{\rm HI}$}
\newcommand{\mhtwo}{$M_{\rm H_2}$}
\newcommand{\mb}{$\rm M_{B}$}
\newcommand{\lfir}{$L_{\rm FIR}$}

\begin{document}

\title{Molecular gas in the central regions of the latest-type spiral galaxies}
\titlerunning{CO in late-type spirals}
\authorrunning{B\"oker \ea }

\author{Torsten B\"oker\inst{1}\fnmsep\thanks{On assignment from the 
Space Telescope Division of the European Space Agency (ESA)} 
\and Ute Lisenfeld\inst{2} 
\and Eva Schinnerer\inst{3}\fnmsep\thanks{Jansky-Fellow}
}   

\offprints{T. B\"oker}

\institute{
Space Telescope Science Institute, 3700 San Martin Drive, Baltimore, MD 21218, U.S.A. \\
\email{boeker@stsci.edu}
\and
Instituto de Astrof\'\i sica de Andaluc\'\i a (CSIC), Camino Bajo de Huetor 24, 18080 Granada, Spain \\
\email{ute@iaa.es}
\and
National Radio Astronomy Observatory, P.O. Box 0, Socorro, NM 87801, U.S.A. \\
\email{eschinne@nrao.edu}
}

\date{Received 11 March 2003; accepted 15 May 2003}

\abstract{
Using the IRAM $30\m$ telescope, we have surveyed an unbiased sample of 
47 nearby spiral galaxies of very late (Scd-Sm) Hubble-type for emission 
in the \coone\ and \two\ lines. The sensitivity of our data (a few mK) 
allows detection of about 60\% of our sample in at least one of the CO 
lines. The median detected \htwo\ mass is $1.4\times 10^7 \msun$ within the 
central few kpc, assuming a standard conversion factor.
We use the measured line intensities to complement existing studies 
of the molecular gas content of spiral galaxies as a function of 
Hubble-type and to significantly improve the statistical significance 
of such studies at the late end of the spiral sequence. 
We find that 
the latest-type spirals closely follow the correlation between molecular 
gas content and galaxy luminosity established for earlier Hubble types.
The molecular gas in late-type galaxies seems to
be less centrally concentrated than in earlier types.
We use {\it Hubble Space Telescope} optical images to correlate the 
molecular gas mass to the properties of the central galaxy disk and the 
compact star cluster that occupies the nucleus of most late-type spirals. 
There is no clear correlation between the luminosity of the nuclear star cluster 
and the molecular gas mass, although the CO detection rate is highest for 
the brightest clusters. It appears that the central surface brightness of the 
stellar disk is an important parameter for the amount of molecular gas at 
the galaxy center. Whether stellar bars play a critical role for 
the gas dynamics remains unclear, in part because of uncertainties in the 
morphological classifications of our sample.
}

\maketitle

\section{Introduction}
Spiral galaxies of the latest Hubble types (between Scd and Sm)
are considered dynamically simple systems: they are ``pure'' disk 
galaxies, often with only weak spiral arm structure, ill-defined 
stellar bars, and no obvious central bulges. Their stellar disks often 
are extremely
thin, as evidenced by a large ratio of radial to vertical disk scale 
lengths in edge-on images \citep{dal00}. In many cases the thin disks
are embedded in a red stellar envelope that most plausibly formed
before the thin disk \citep{dal02}. The undisturbed morphology
of these systems indicates that they have not experienced any
significant merger events after the formation of the red envelopes
($\approx 6\gyr$ ago). Late-type spirals thus provide important
constraints on galaxy formation scenarios that invoke hierarchical
merging.

In addition, the featureless stellar disks, shallow
\sbp s, and lack of bulge-like structures apparent in optical
images of many such galaxies \citep{mat97,boe03} suggest 
an uneventful star formation history within their central few kpc. 
It is unclear whether this lack of past star formation activity
in the central region is simply due to a lack of cold molecular 
gas (which provides 
the raw material for star formation), or whether dynamical effects 
prevent the gas from reaching a critical density. Unfortunately, late-type
spirals are under-represented in many samples because of observational 
biases. This is especially true for surveys of their molecular gas content.
For example, a literature compilation of the CO emission 
in 582 galaxies by \cite{cas98} contains only 55 spirals with Hubble type
Scd or later, only 26 of which have actual detections. With the data
presented in this paper, we more than double the amount of available
CO observations of late-type spirals.
More specifically, we present spectra of the $\rm ^{12}CO$ emission 
from 47 objects in both the \one\ and \two\ lines, 30 of which 
have reliable detections in at least one of the CO lines.

Our study is further motivated by an apparent conundrum that has
recently emerged from high-resolution images of this galaxy class.
Over the last few years, {\it Hubble Space Telescope (HST)} images
have revealed that the photocenter of many late-type spirals is 
occupied by a compact, luminous stellar cluster \citep{phi96,car98,mat99}. 
In a recent paper \citep[][hereafter paper I]{boe02}, we have shown that at 
least 75\% of 
spiral galaxies with Hubble types later than Sc harbor such a nuclear star 
cluster. It thus appears that even in the most shallow disks, the nucleus 
is a well-defined location. This is difficult to explain, given the fact 
that the slowly rising rotation curves observed in most late-type
spirals indicate a nearly homogenous mass distribution over much of the 
central disk \citep[e.g.][]{mat97}, and hence gravity does not
provide a strong force towards the nucleus. 

Even more surprising is the fact that many of the \nc s for which
spectroscopic information exists are dominated by a relatively young 
($\approx 100\myr$) stellar population \citep{gor99,boe99,boe01,dav02}. 
The dynamical masses for 
these nuclear star clusters are in the range $5\cdot 10^5 - 10^7\msun$ 
\citep[][]{kor93,boe99,mat02,wal03}, and it is unclear how the high
gas densities required for the formation of such dense and massive 
star clusters could have been achieved in the centers of late-type spirals.
One possible scenario is that the clusters are formed over time
by a series of modest starburst events, rather than in a single,
massive collapse. This model is difficult to confirm observationally,
because the cluster spectrum is always dominated by the youngest 
stellar population which contains the most massive and brightest stars.
The purpose of this paper is to provide a ``sanity check'' of this
scenario by quantifying the amount of molecular material in the central 
few kpc of late-type spirals. Clearly, if repetitive starbursts
are a viable model for nuclear cluster formation, one would expect
to find a central reservoir of molecular gas from which gas can be 
transported into the central few pc. 

In Sect.~\ref{sec:obs}, we describe our galaxy sample, the details
of the CO observations, and the data reduction procedure. The resulting
CO spectra and their quantitative analysis are presented in
Sect.~\ref{sec:results}. In Sect.~\ref{sec:analysis}, we compare the results
for our sample to those derived for early-type disk galaxies. In addition, 
we use the HST images of paper~I to investigate possible dependencies 
between the molecular gas content and galaxy parameters such as 
luminosity, bar class, or surface brightness, as well as the luminosity 
of the nuclear star clusters. We summarize and conclude in Sect.~\ref{sec:summary}.
%
\section{Sample Selection, Observations and Data Reduction} \label{sec:obs}
\subsection{The galaxy sample}\label{subsec:sample}
Our galaxy sample was selected according to the criteria described
in paper\,I: 
\ben
\item{Hubble type between Scd and Sm ($ 6 \leq T \leq 9$).}
\item{Line-of-sight velocity $v_{\rm hel} < 2000\kms$.}
\item{Axis ratio parameter $R_{25} \equiv {\rm log}(a/b) < 0.2$, i.e. 
inclination close to face-on.}
\een
The {\it NASA Extragalactic Database, (NED)} contains 120 galaxies 
that meet the above criteria. In this paper, we present new IRAM 
observations for a subset of 47 objects. The galaxy names and some 
selected properties of our IRAM sample are listed in Table~\ref{tab:obs}.

%
\begin{table*}
\scriptsize
\caption{Galaxy sample \label{tab:obs}}
\begin{tabular}{lccccccccc}
\hline
(1) & (2) & (3) & (4) & (5) & (6) & (7) & (8) & (9) & (10) \\
Galaxy & R.A. & Dec. & $d$ & Type & $m_{\rm B}$ & $W_{20}$ & 
$d_{25}$ & $\rm \mu_I^0$ & $\rm M_I^{cl}$ \\
 & (J2000) & (J2000) & [Mpc] &  & [mag] & [$\kms$] & [arcmin] & [$\rm mag\,arcsec^{-2}$] \\
\hline
NGC\,337a   & 01 01 33.90 & -07 35 17.7 &  14.3  & SAB(s)dm  & 13.38 & $ 98\pm 7 $ & 5.0 & 19.9 & -10.02  \\
MCG\,1-3-85 & 01 05 04.88 & -06 12 45.9 &  14.6  & SAB(rs)d  & 12.62 & $182\pm 7 $ & 4.3 & na	&   na  \\
NGC\,428    & 01 12 55.60 & -00 58 54.4 &  16.1  & SAB(s)m   & 12.03 & $179\pm 8 $ & 3.6 & 18.7 & -13.15  \\
UGC\,3574   & 06 53 10.60 & +57 10 39.0 &  23.4  & SA(s)cd   & 14.59 & $159\pm 6 $ & 3.4 & 18.4 & -11.90  \\
UGC\,3826   & 07 24 32.05 & +61 41 35.2 &  27.8  & SAB(s)d   & 15.00 & $ 65\pm 6 $ & 3.5 & 19.1 & -10.76  \\
NGC\,2552   & 08 19 20.14 & +50 00 25.2 &   9.9  & SA(s)m?   & 12.81 & $144\pm 5 $ & 3.4 & 19.7 & -12.04  \\
UGC\,4499   & 08 37 41.43 & +51 39 11.1 &  12.5  & SAdm      & 14.82 & $145\pm 19$ & 2.4 & 19.8 & -8.59  \\
NGC\,2805   & 09 20 24.56 & +64 05 55.2 &  28.1  & SAB(rs)d  & 11.93 & $122\pm 13$ & 5.9 & 18.0 & -13.32  \\
UGC\,5015   & 09 25 47.89 & +34 16 35.9 &  25.7  & SABdm     & 15.60 & $418\pm 14$ & 1.8 & 19.1 & -11.37  \\
UGC\,5288   & 09 51 17.00 & +07 49 39.0 &   8.0  & Sdm:      & 15.55 & $108\pm 6 $ & 0.8 & 19.8 &   nc  \\
NGC\,3206   & 10 21 47.65 & +56 55 49.6 &  19.7  & SB(s)cd   & 13.94 & $194\pm 13$ & 2.8 & 18.8 &   nc  \\
NGC\,3346   & 10 43 38.90 & +14 52 18.0 &  18.8  & SB(rs)cd  & 12.72 & $166\pm 11$ & 2.7 & 18.1 & -11.78  \\
NGC\,3423   & 10 51 14.30 & +05 50 24.0 &  14.6  & SA(s)cd   & 11.59 & $184\pm 11$ & 3.9 & 17.4 & -11.84  \\
NGC\,3445   & 10 54 35.87 & +56 59 24.4 &  32.1  & SAB(s)m   & 12.90 & $164\pm 14$ & 1.5 & 17.6 & -13.42  \\
NGC\,3782   & 11 39 20.72 & +46 30 48.6 &  13.5  & SAB(s)cd: & 13.13 & $133\pm 8 $ & 1.6 & 18.2 & -10.07  \\
NGC\,3906   & 11 49 40.46 & +48 25 33.3 &  16.7  & SB(s)d    & 13.72 & $ 48\pm 7 $ & 1.6 & 18.6 & -10.01  \\
NGC\,3913   & 11 50 38.77 & +55 21 12.1 &  17.0  & SA(rs)d:  & 13.37 & $ 54\pm 6 $ & 2.4 & 17.6 & -9.96  \\
UGC\,6931   & 11 57 22.79 & +57 55 22.5 &  20.7  & SBm:      & 15.14 & $127\pm 8 $ & 1.3 & 20.1 & -9.72  \\
NGC\,4204   & 12 15 14.51 & +20 39 30.7 &  13.8  & SB(s)dm   & 14.01 & $ 99\pm 12$ & 3.8 & 19.4 & -10.26  \\
NGC\,4242   & 12 17 30.10 & +45 37 07.5 &  10.5  & SAB(s)dm  & 11.69 & $137\pm 8 $ & 4.8 & na	& -11.33$^*$  \\
NGC\,4299   & 12 21 40.90 & +11 30 03.0 &  16.8  & SAB(s)dm: & 13.01 & $148\pm 12$ & 1.6 & 17.7 & -11.73  \\
NGC\,4395   & 12 25 48.92 & +33 32 48.4	&   7.1  & SA(s)m    & 11.40 & $131\pm 7 $ & 12.2 & na  & -12.33$^*$  \\
NGC\,4416   & 12 26 46.72 & +07 55 07.9 &  20.7  & SB(rs)cd: & 13.24 & $159\pm 16$ & 1.6 & 17.9 & -8.81  \\
NGC\,4411B  & 12 26 47.30 & +08 53 04.5 &  19.1  & SAB(s)cd  & 13.24 & $ 95\pm 16$ & 2.4 & 16.9 & -12.57 \\
NGC\,4487   & 12 31 04.36 & -08 03 13.8 &  14.6  & SAB(rs)cd & 12.21 & $224\pm 8 $ & 3.7 & 17.0 & -12.97  \\
NGC\,4496A  & 12 31 39.32 & +03 56 22.7 &  25.3  & SB(rs)m   & 12.12 & $175\pm 7 $ & 3.8 & 18.6 & -11.99 \\
NGC\,4517A  & 12 32 28.15 & +00 23 22.8 &  22.2  & SB(rs)dm: & 13.19 & $164\pm 4 $ & 3.8 & 19.8 &   nc \\
NGC\,4519   & 12 33 30.27 & +08 39 17.0 &  18.5  & SB(rs)d   & 12.50 & $209\pm 13$ & 2.9 & na	&   na \\
NGC\,4534   & 12 34 05.44 & +35 31 08.0 &  14.2  & SA(s)dm:  & 13.04 & $127\pm 19$ & 2.9 & na	&   na \\
NGC\,4540   & 12 34 50.90 & +15 33 06.9 &  19.8  & SAB(rs)cd & 12.54 & $178\pm 42$ & 2.1 & 17.6 & -12.29  \\
NGC\,4618   & 12 41 32.74 & +41 09 03.8 &  10.7  & SB(rs)m   & 11.45 & $131\pm 28$ & 4.3 & 18.0 & -11.45  \\
NGC\,4625   & 12 41 52.61 & +41 16 26.3 & 11.7& SAB(rs)m pec & 13.08 & $ 73\pm 7 $ & 2.2 & 16.8 & -10.61  \\
NGC\,4688   & 12 47 46.77 & +04 20 08.8 &  14.9  & SB(s)cd   & 13.52 & $ 71\pm 6 $ & 3.7 & na	&   na  \\
NGC\,4701   & 12 49 11.71 & +03 23 21.8 &  11.0  & SA(s)cd   & 12.91 & $174\pm 10$ & 2.6 & 15.9 & -13.45  \\
NGC\,4775   & 12 53 45.79 & -06 37 20.1 &  22.4  & SA(s)d    & 12.20 & $123\pm 13$ & 2.2 & 16.8 & -13.77  \\
UGC\,8516   & 13 31 52.50 & +20 00 01.0 &  16.5  & Scd:      & 14.37 & $121\pm 6 $ & 1.1 & 18.1 & -10.97  \\
NGC\,5477   & 14 05 31.25 & +54 27 12.3 &   8.1  & SA(s)m    & 14.62 & $ 70\pm 10$ & 1.4 & 20.4 &   nc   \\
NGC\,5584   & 14 22 23.65 & -00 23 09.2 &  24.2  & SAB(rs)cd & 12.51 & $211\pm 6 $ & 3.3 & 17.7 & -9.47  \\
NGC\,5669   & 14 32 44.00 & +09 53 31.0	&  21.2  & SAB(rs)cd & 13.11 & $213\pm 9 $ & 4.1 & 18.5 & -10.03  \\
NGC\,5668   & 14 33 24.30 & +04 27 02.0 &  23.8  & SA(s)d    & 12.51 & $115\pm 11$ & 2.8 & 17.8 & -13.10  \\
NGC\,5725   & 14 40 58.30 & +02 11 10.0 &  24.4  & SB(s)d:   & 14.73 & $157\pm 7 $ & 1.0 & na	&   na  \\
NGC\,5789   & 14 56 35.52 & +30 14 02.5 &  28.6  & Sdm       & 14.44 & $157\pm 12$ & 1.1 & 19.9 &   nc  \\
NGC\,5964   & 15 37 36.30 & +05 58 26.0 &  22.2  & SB(rs)d   & 13.23 & $192\pm 9 $ & 4.1 & 18.4 & -12.62  \\
NGC\,6509   & 17 59 25.36 & +06 17 12.4 &  27.5  & SBcd      & 13.37 & $276\pm 20$ & 1.5 & 17.4 & -13.08  \\
UGC\,12082  & 22 34 11.54 & +32 52 10.3 &  13.9  & Sm	     & 14.14 & $ 79\pm 6 $ & 2.8 & 20.7 &   nc  \\
UGC\,12732  & 23 40.39.80 & +26 14 10.0 &  12.4  & Sm:	     & 14.26 & $131\pm 10$ & 3.3 & 20.1 & -11.29  \\
NGC\,7741   & 23 43 53.65 & +26 04 33.1 &  12.5  & SB(s)cd   & 12.09 & $209\pm 8 $ & 4.1 & 18.6 &   nc \\
\hline
\end{tabular}

Columns 1-3: Galaxy name and coordinates. Column\,4:
Galaxy distance (in Mpc), calculated from the measured recession velocities, 
corrected for Virgo-centric infall, and assuming $\rm H_0 = 70\kms\,Mpc^{-1}$.
Column\,5: morphological type as listed in the RC3. Columns 6-8: galaxy
total apparent blue magnitude, $21\cm$ line width (measured at the 
20\% level), and apparent optical diameter as listed in LEDA. Column\,9: central 
I-band surface brightness of the stellar disk, derived from the analysis of HST 
images described in paper~I. Galaxies marked `na' are not in the sample of paper~1. 
Column\,10: absolute I-band magnitude of the nuclear star cluster, taken from paper~1
except for NGC\,4242 and NGC\,4395 (flagged by a $*$) which are taken from \cite{mat99}.
Galaxies in which no nuclear cluster could be identified are marked `nc'. 

\end{table*}
We point out that the sample selection criteria are 
unbiased (within the limits of the catalogs that form the basis
of {\it NED}) with respect to galaxy size, stellar or gaseous mass, 
total magnitude, star formation efficiency, or any other quantity 
that might reasonably be expected to favor or disfavor nuclear
star formation. It should therefore be well suited to provide 
a representative measure of the molecular gas content in 
late-type galaxies in the local universe.

Of the 47 IRAM targets, all but 7 (MCG-1-3-85, NGC\,4242, NGC\,4395,
NGC\,4519, NGC\,4534, NGC\,4688, and NGC\,5725) have HST I-band 
observations that were presented in paper\,I. We use these images 
to search for possible correlations of the molecular gas content both 
with the central surface brightness of the stellar disk and the 
luminosity of the nuclear star cluster (Sect.~\ref{subsec:disk} and 
\ref{subsec:cluster}). The spatial resolution afforded by HST is 
essential for this purpose because the emission from the luminous 
central star cluster is difficult to separate from the underlying 
disk, even in the best seeing conditions.
%
\subsection{Observations and data reduction}
All objects were observed with the IRAM $30\m$ telescope on Pico Veleta,
Spain between December 2001 and August 2002.  
We used dual polarization receivers at the (redshifted) frequencies of the 
\coone\ and \cotwo\ lines at 115 and $230\ghz$, respectively, employing the
$512 \times 1\mhz$ filterbanks for the \one\ line and the autocorrelator
for the \two\ line. All observations were done in wobbler switching mode
with a wobbler throw of $200\as$ in azimuth. The galaxies were observed at 
their central positions listed in Table~\ref{tab:obs}, with one beam position
per galaxy. The telescope pointing accuracy was monitored on nearby quasars 
every 60 -- 90 minutes, the rms offset being $\approx 4-5\as$. 
Due to relatively
poor weather conditions and anomalous refraction during most of our
observations, the pointing accuracy is less than optimal. However, given the
half-power beam width (HPBW) of the $30\m$ telescope of 
$21\as$ at $115\ghz$ and 
$11\as$ at $230\ghz$, this is adequate to measure the nuclear CO emission. 
At the median distance of our galaxy sample ($16.7\mpc$), the beam size 
corresponds to $1.7\kpc$ ($0.9\kpc$) at $115\ghz$ ($230\ghz$). 

Typical on-source integration times were 0.5--1 hours per object, 
divided in scans of 6 minutes. The individual scans were averaged 
after flagging of bad channels, and linear baselines were subtracted. 
Typical rms noise levels were in the range $\rm 3-10\>mK$ for the 
\one\ line and $\rm 5-38\>mK$ for the \two\ line (after smoothing
to a velocity resolution of $10\kms$). The reduced sensitivity of some
of the \two\ observations is due to the fact that during part of the
observations, only one of the two receivers was functional. 

The data are calibrated to the scale of corrected antenna temperature, 
$T_A^*$,  by measuring loads at ambient and cold temperature, as in the 
conventional ``chopper wheel'' calibration for millimeter wavelength observations. 
A calibration measurement was carried out every 15-20 minutes during an
integration and every time a new object was acquired. 
Typical system temperatures were in the range $200-300\kk$ at $115\ghz$
and $250-600\kk$ at $230\ghz$ on the $T_A^*$ scale. 
The IRAM forward efficiency, $F_{\rm eff}$, was 0.95 and 0.91 at 115 
and $230\ghz$ and the beam efficiency, $B_{\rm eff}$, was 0.75 and 0.50, 
respectively. All CO spectra and luminosities are
presented in the main beam temperature scale ($T_{\rm mb}$) which is
defined as $T_{\rm mb} = (F_{\rm eff}/B_{\rm eff})\times T_A^*$.
The conversion factor from main beam temperature to flux densities
for the $30\m$ telescope is $\rm 4.8\>Jy/K$. 
\section{Results} \label{sec:results}
\subsection{CO spectra} \label{subsec:spectra}
\begin{figure*}
\centering
\includegraphics[width=17cm]{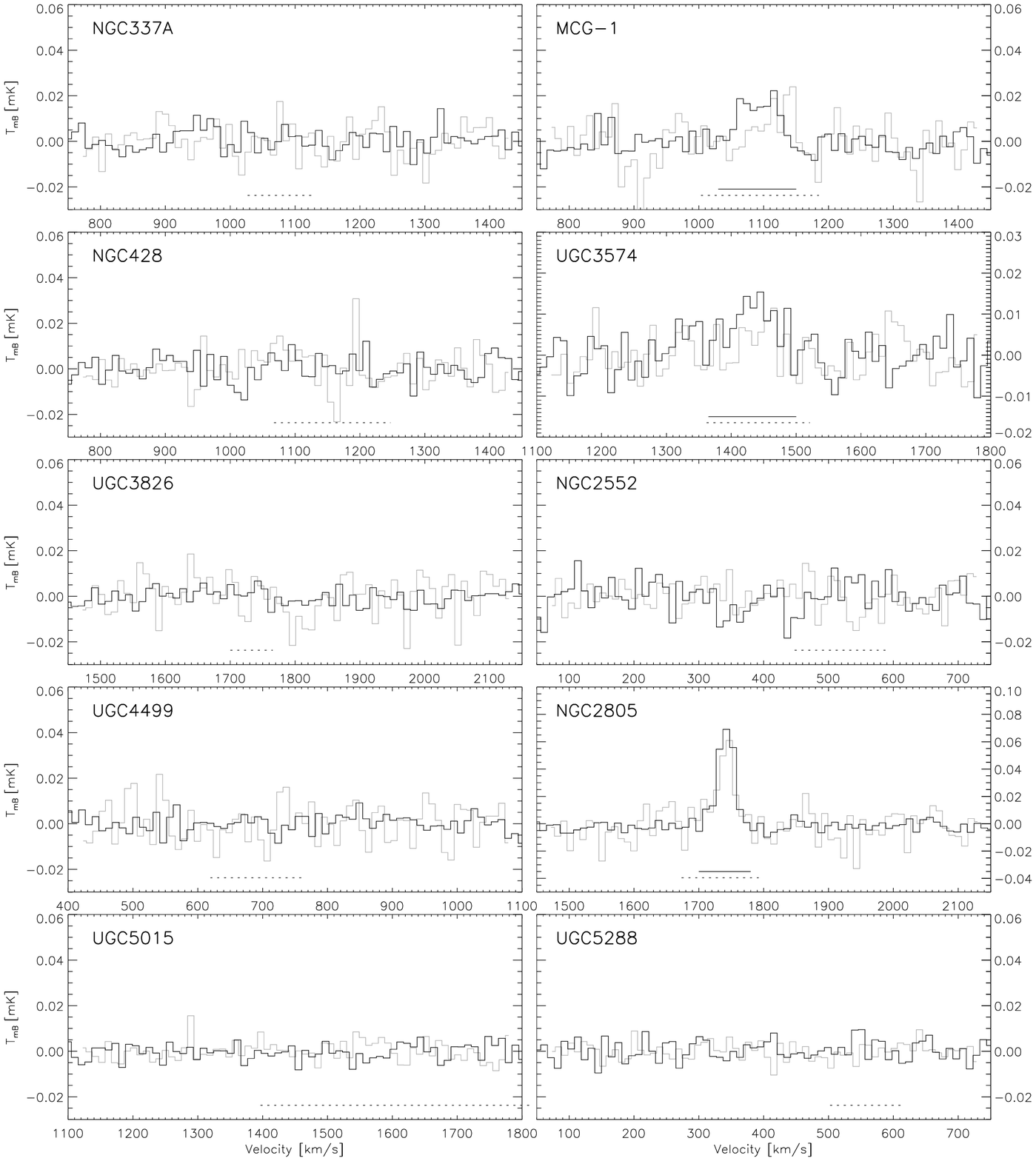}
\caption{\label{fig:spectra}
  Calibrated \coone\ (black) and \cotwo\ (grey) spectra for all 47
  galaxies. The spectra have been smoothed to a velocity resolution
  of $10\kms$. For those galaxies which are detected
  in at least one CO line, the solid horizontal line indicates
  the spectral window over which the CO line fluxes
  have been integrated. For comparison, the 
  dashed line in each panel denotes the width of the $21\cm$ line, 
  measured at the 20\% level (Col. 7 of Table~\ref{tab:obs})
  and centered on the observed galaxy recession velocity. 
  Both central velocity and width of the $21\cm$ line 
  are taken from the LEDA database.
}
\end{figure*}

In Fig.~\ref{fig:spectra}, we show the calibrated \coone\ and \two\ 
spectra of all observed galaxies. The spectra were smoothed to a
velocity resolution of $\sim 10\kms$. We detect 29 galaxies in the 
\coone\ line, 21 of which also show evidence for \two\ emission. 
One object (NGC~5584) is detected only in \two .
The \coone\ detections for most galaxies are more than 5$\sigma$ above
the noise, except for NGC\,3906, NGC\,4204 and NGC\,4299 for which
the detections are only at a level of $3\sigma$.

\begin{table*}
\scriptsize
\caption{Results of Observations \label{tab:res}}
\begin{tabular}{lccccccc}
\hline
(1) & (2) & (3) & (4) & (5) & (6) & (7) & (8)  \\
Galaxy & \ione & \itwo & $W_{50}^{\rm CO}$ & $v_{\rm CO}$ & 
\mhtwo & \mhtwo (var) & \mhi \\ 
 & [$\kkms$] & [$\kkms$]     & [$\kms$] & 
     [$\kms$]  & [$10^6\msun$] & [$10^6\msun$] & [$10^9\msun$] \\
\hline
NGC\,337a   &	   $<$   0.46 &    $<$ 1.01	&  --	     &  --  &  $<$  4.0  & $<$ 14.4 & 4.6 \\
MCG\,1-3-85 & 1.39 $\pm$ 0.18 & 1.39 $\pm$ 0.49 & 67$\pm$8   & 1092 &	   12.7  &     22.2 & 3.6 \\
NGC\,428    &	     $<$ 0.74 &    $<$ 1.23	&  --	     &   -- &  $<$  8.2  & $<$  7.6 & 5.0 \\
UGC\,3574   & 1.16 $\pm$ 0.18 &    $<$ 0.78	& 123$\pm$32 & 1426 &	   27.2  &     30.7 & 5.8 \\
UGC\,3826   &	     $<$ 0.26 &    $<$ 0.98	&  --	     &   -- &  $<$  8.6  & $<$ 12.1 & 5.0 \\
NGC\,2552   &	     $<$ 0.79 &    $<$ 0.99	&  --	     &   -- &  $<$  3.3  & $<$  6.2 & 0.7 \\
UGC\,4499   &	     $<$ 0.44 &    $<$ 1.53	&  --	     &   -- &  $<$  2.9  & $<$  6.6 & 1.1 \\
NGC\,2805   & 2.34 $\pm$ 0.10 & 3.28 $\pm$ 0.48 & 29$\pm$1   & 1742 &	   79.2  &     37.7 & 18.5 \\
UGC\,5015   &	     $<$ 0.71 &    $<$ 0.95	&  --	     &   -- &  $<$ 20.1  & $<$ 42.1 & 1.3 \\
UGC\,5288   &	     $<$ 0.53 &    $<$ 0.47	&  --	     &   -- &  $<$  1.5  & $<$  6.2 & 0.3 \\
NGC\,3206   &	     $<$ 0.52 &    $<$ 1.08	&  --	     &   -- &  $<$  8.6  & $<$  8.8 & 3.2 \\
NGC\,3346   & 3.58 $\pm$ 0.23 & 1.57 $\pm$ 0.66 &  81$\pm$6  & 1244 &	   54.2  &     53.6 & 1.4 \\
NGC\,3423   & 2.70 $\pm$ 0.24 & 2.67 $\pm$ 0.70 &  66$\pm$6  & 1002 &	   24.7  &     21.8 & 2.3 \\
NGC\,3445   & 0.89 $\pm$ 0.10 & 1.41 $\pm$ 0.46 &  40$\pm$5  & 2042 &	   39.3  &     29.4 & 4.7 \\
NGC\,3782   & 0.68 $\pm$ 0.11 & 1.05 $\pm$ 0.33 & 56$\pm$10  &  753 &	    5.3  &      9.4 & 1.3 \\
NGC\,3906   & 0.37 $\pm$ 0.12 &    $<$ 0.86	&  19$\pm$6  &  969 &	    4.4  &      7.6 & 0.3 \\
NGC\,3913   & 1.88 $\pm$ 0.22 & 1.98 $\pm$ 0.67 &  24$\pm$2  &  961 &	   23.3  &     34.6 & 0.9 \\
UGC\,6931   &	     $<$ 0.50 &    $<$ 1.23	&  --	     &   -- &  $<$  9.2  & $<$ 18.3 & 0.6 \\
NGC\,4204   & 0.50 $\pm$ 0.14 &    $<$ 1.65	&  34$\pm$13 &  849 &	      4.1  &	6.5 & 1.5 \\
NGC\,4242   &	     $<$ 0.90 &    $<$ 0.98	&  --	     &   -- &  $<$  4.3  & $<$  4.6 & 1.2 \\
NGC\,4299   & 0.48 $\pm$ 0.14 & 0.54 $\pm$ 0.19 &  24$\pm$8  &  223 &	      5.8  &	7.7 & 1.1 \\
NGC\,4395   &	     $<$ 0.58 &     $<$ 0.70	&  --	     &   -- &  $<$  1.3  & $<$  1.4 & 3.7 \\
NGC\,4416   & 5.08 $\pm$ 0.19 & 4.58 $\pm$ 0.34 &  78$\pm$3  & 1391 &	   93.3  &    114.5 & 0.5 \\
NGC\,4411B  & 0.88 $\pm$ 0.10 &    $<$ 0.56	&  37$\pm$5  & 1268 &	   13.8  &     16.5 & 1.2 \\
NGC\,4487   & 3.23 $\pm$ 0.15 & 3.74 $\pm$ 0.55 & 80 $\pm$4  & 1026 &	   29.5  &     34.4 & 1.8 \\
NGC\,4496A  & 2.17 $\pm$ 0.18 & 2.23 $\pm$ 0.63 &  48$\pm$5  & 1734 &	   59.5  &     37.1 & 7.2 \\
NGC\,4517A  &	     $<$ 0.65 &    $<$ 1.03	&  --	     &   -- &  $<$ 13.7  & $<$ 14.6 & 5.1 \\
NGC\,4519   & 2.93 $\pm$ 0.27 & 2.31 $\pm$ 0.77 &  92$\pm$10 & 1231 &	   43.0  &     41.9 & 3.9 \\
NGC\,4534   &	     $<$ 0.42 &    $<$ 1.31	&  --	     &   -- &  $<$  3.6  & $<$  5.7 & 3.1 \\
NGC\,4540   & 5.57 $\pm$ 0.19 & 5.25 $\pm$ 0.39 &  75$\pm$3  & 1303 &	   93.6  &     89.5 & 0.6 \\
NGC\,4618   & 0.72 $\pm$ 0.17 & 0.80 $\pm$ 0.13 &  19$\pm$5  &  534 &	      3.5  &	3.5 & 2.3 \\
NGC\,4625   & 3.78 $\pm$ 0.11 & 4.32 $\pm$ 0.46 &  37$\pm$1  &  615 &	   22.2  &     41.5 & 1.0 \\
NGC\,4688   & 0.44 $\pm$ 0.08 &    $<$ 0.41	&  20$\pm$5  &  983 &	      4.2  &	5.6 & 1.9 \\
NGC\,4701   & 2.90 $\pm$ 0.22 & 3.02 $\pm$ 0.40 &  99$\pm$10 &  704 &	   15.0  &     28.3 & 1.6 \\
NGC\,4775   & 1.63 $\pm$ 0.22 & 1.18 $\pm$ 0.23 &  31$\pm$6  & 1562 &	   35.1  &     27.6 & 3.7 \\
UGC\,8516   & 1.20 $\pm$ 0.19 &    $<$ 1.35	&  69$\pm$13 & 1026 &	   14.0  &     30.5 & 0.3 \\
NGC\,5477   &	     $<$ 0.43 &    $<$ 0.40	&  --	     &   -- & $<$   1.2  &  $<$ 5.7 & 0.2 \\
NGC\,5584   &	     $<$ 0.50 & 2.1$\pm$0.57	&  14$\pm$5$^*$  & 1637$^*$ &	13.2$^*$ & 11.4$^*$ & 4.0 \\
NGC\,5669   & 2.28 $\pm$ 0.27 & 3.02 $\pm$ 0.67 &  86$\pm$15 & 1369 &	   43.9  &     33.3 & 4.7 \\
NGC\,5668   & 1.78 $\pm$ 0.26 & 1.52 $\pm$ 0.66 &  41$\pm$7  & 1602 &	   43.2  &     31.7 & 6.1 \\
NGC\,5725   & 1.12 $\pm$ 0.24 &    $<$ 3.90	&  45$\pm$19 & 1648 &	   28.6  &     47.9 & 0.5 \\
NGC\,5789   &	     $<$ 0.74 &    $<$ 3.90	&  --	     &   -- &  $<$ 25.9  & $<$ 45.4 & 1.4 \\
NGC\,5964   & 0.89 $\pm$ 0.17 &    $<$ 0.67	&  43$\pm$8  & 1471 &	   18.8  &     17.3 & 5.1 \\
NGC\,6509   & 5.97 $\pm$ 0.23 & 4.85 $\pm$ 0.98 & 118$\pm$5  & 1830 &	  193.5  &    180.9 & 6.2 \\
UGC\,12082  &	     $<$ 0.46 &    $<$ 1.76	&  --	     &   -- &  $<$  3.8  & $<$ 10.0 & 1.4 \\
UGC\,12732  &	     $<$ 0.40 &    $<$ 0.94	&  --	     &   -- &  $<$  2.6  & $<$  6.7 & 2.5 \\
NGC\,7741   & 1.63 $\pm$ 0.20 & 1.75 $\pm$ 0.24 & 114$\pm$15 &  767 &	   10.9  &     12.3 & 1.9 \\
\hline
\end{tabular}

Columns\,2 and 3: Velocity integrated intensity of the \coone\ and
\two\ lines. The temperature refers to the main beam temperature 
scale ($T_{\rm mb}$).
Upper limits were derived as discussed in Sect.~\ref{sec:results}.
Column\,4 and 5: FWHM and central velocity of the \one\ line derived from 
Gauss-fits to the line, except in the case of NGC\,7741 (see text). 
Columns\,6 and 7: Molecular gas mass derived with a constant (6) and
luminosity-dependent (7) conversion factor (see Sect.~\ref{subsec:gasmass}).  
The values for NGC\,5584 (flagged with a $*$) were derived
from the \two\ line intensity.
Column~8: Atomic gas mass, calculated from the $21\cm$ flux.
See Sect.~\ref{subsec:gasmass} for details.

\end{table*}
The quantitative analysis of the spectra is summarized in Table~\ref{tab:res}.
For all detected galaxies, we list widths (at the 50\% level) and 
central velocities of the \one\ emission as well as the integrated 
line intensities. 
Except for one case, the galaxies in our sample show no clear evidence for 
double-horn profiles. The CO lines are well-fitted by single Gaussians;
both widths and central velocities listed in Table~\ref{tab:res} are
derived from these fits. The only exception is NGC\,7741 for which
these numbers were determined from the spectrum itself.
For non-detections, we report upper limits for the CO intensities which
were derived according to
\beq
I_{\rm CO} = 3\sigma \sqrt{\Delta V_{\rm CO}\,\delta V} \>\>\>\> [\kkms] .
\eeq
Here, $\sigma$ is the rms noise of the spectrum, $\delta V$ the 
spectral resolution, and $\Delta V_{\rm CO}$ the total width of the 
detected CO line as indicated by the solid horizontal line in the 
spectra of Fig.~\ref{fig:spectra}. 
In cases for which no CO line was detected, we substitute $\Delta V_{\rm CO}$ with
the width of the HI $21\cm$ line, measured at the 20\% level (Col.~7
in Table~\ref{tab:obs}, dashed horizontal lines in Fig.~\ref{fig:spectra}).

For detections, both line intensities were integrated over the 
\one\ line width, except for NGC\,5584 which was only detected in the
\two\ line. 
In general, the CO line positions agree well with those of the $21\cm$ line.
The errors for the line intensities were calculated as
\beq
\delta I_{\rm CO} = \sigma \sqrt{\Delta V_{\rm CO}\,\delta V} \>\>\>\> [\kkms].
\eeq

\clearpage

\includegraphics[width=17cm]{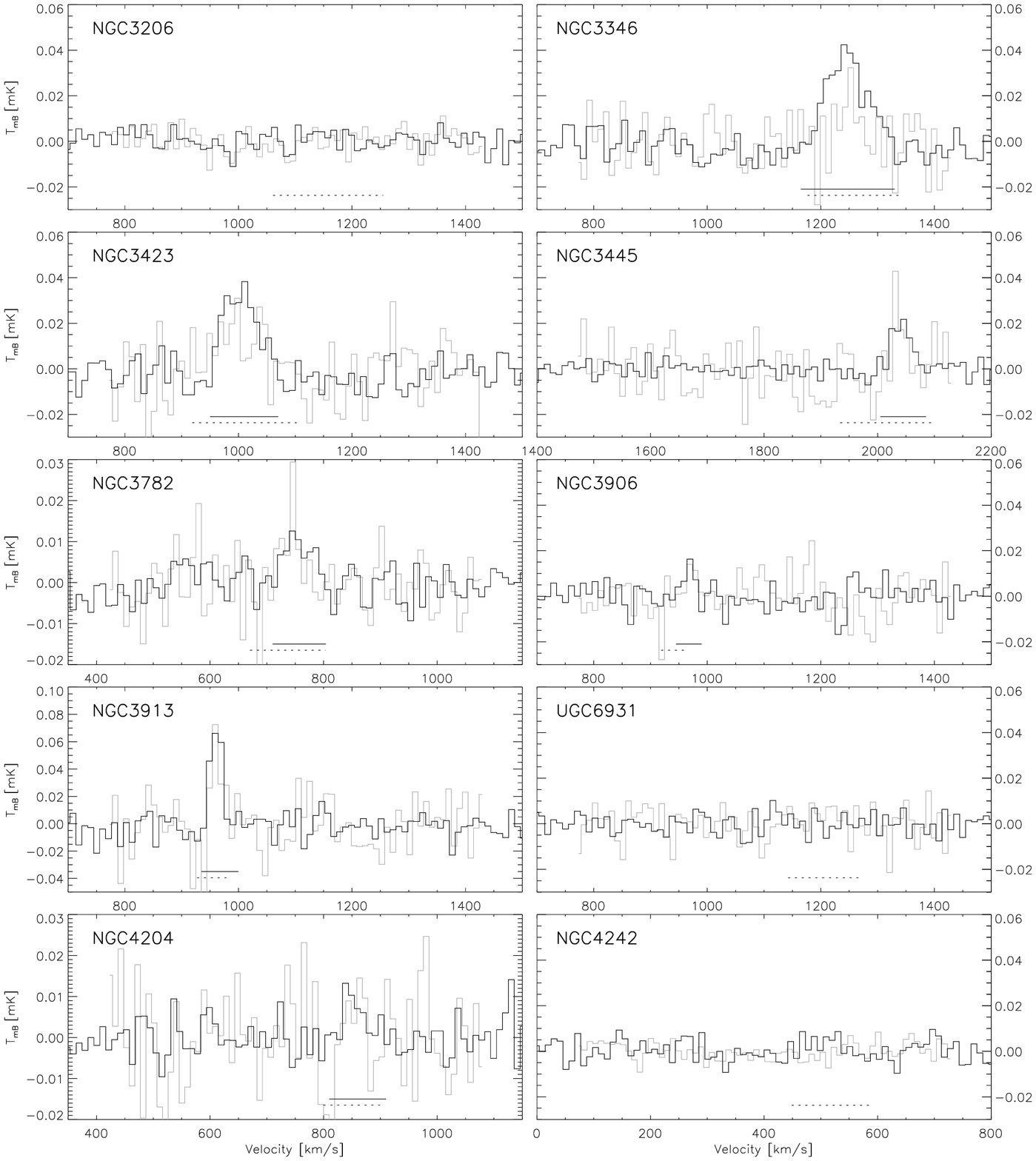}
\centerline{Fig.1 (cont.)}

\clearpage
\includegraphics[width=17cm]{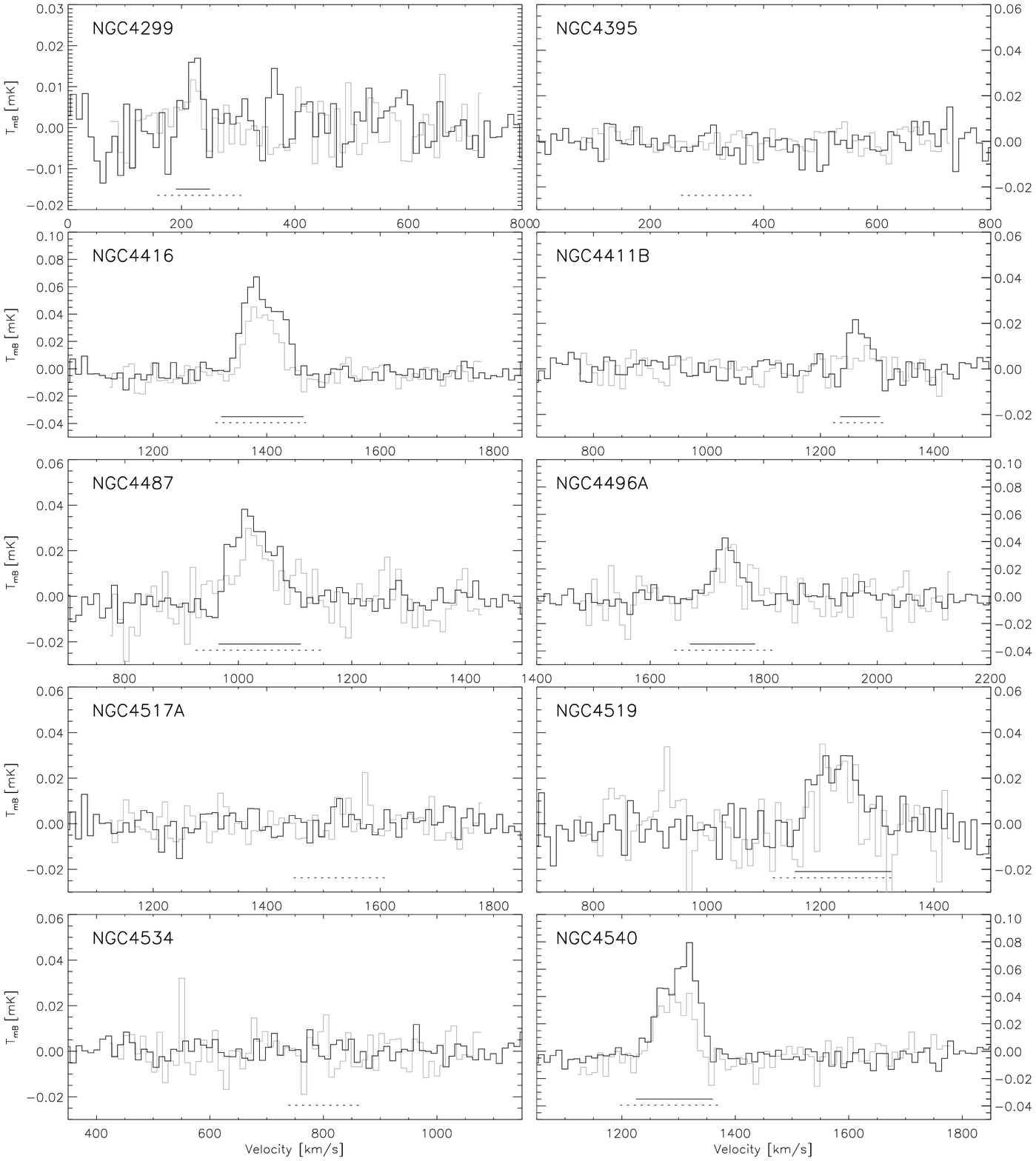}
\centerline{Fig.1 (cont.)}

\clearpage
\includegraphics[width=17cm]{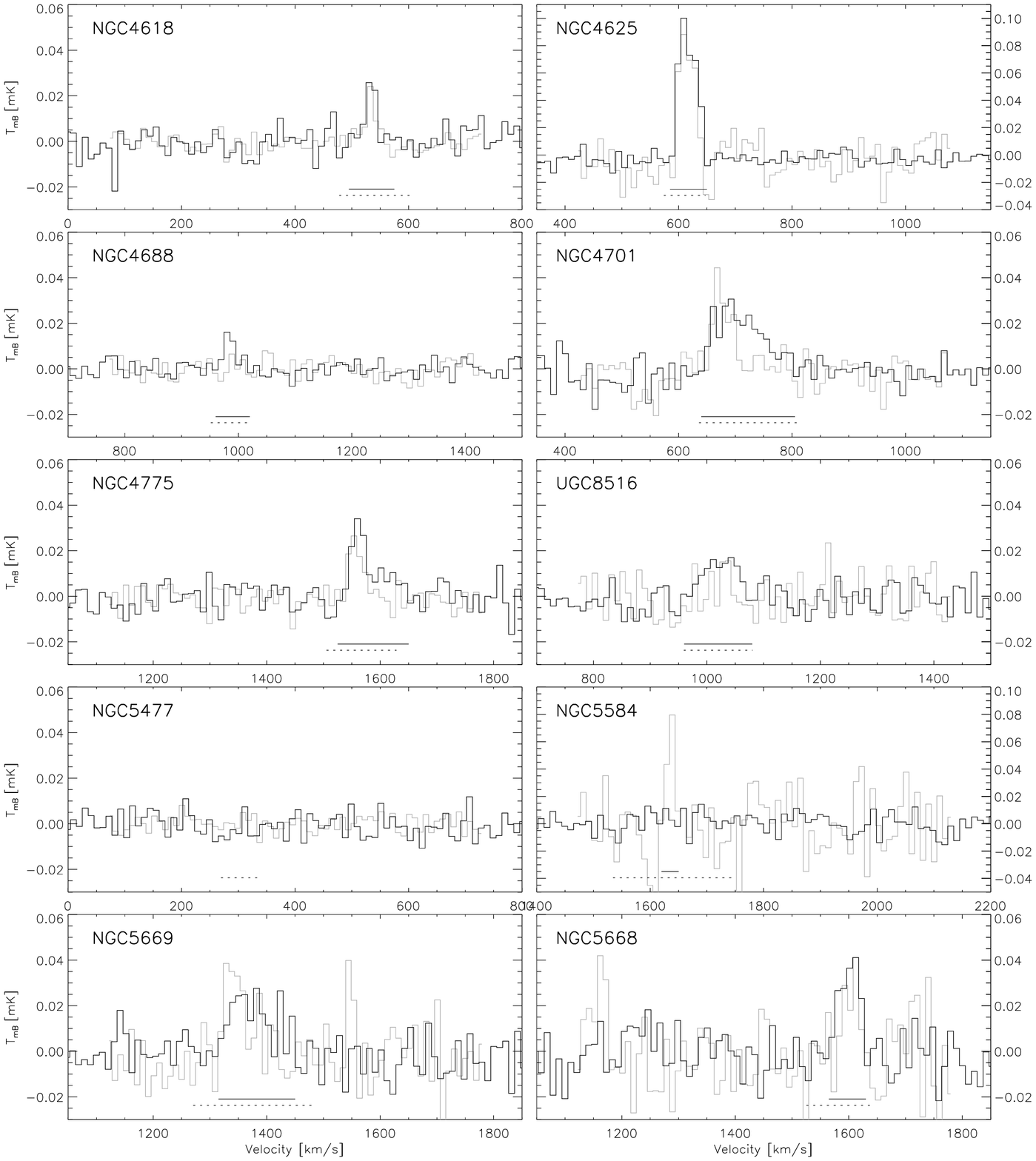}
\centerline{Fig.1 (cont.)}

\clearpage
\includegraphics[width=17cm]{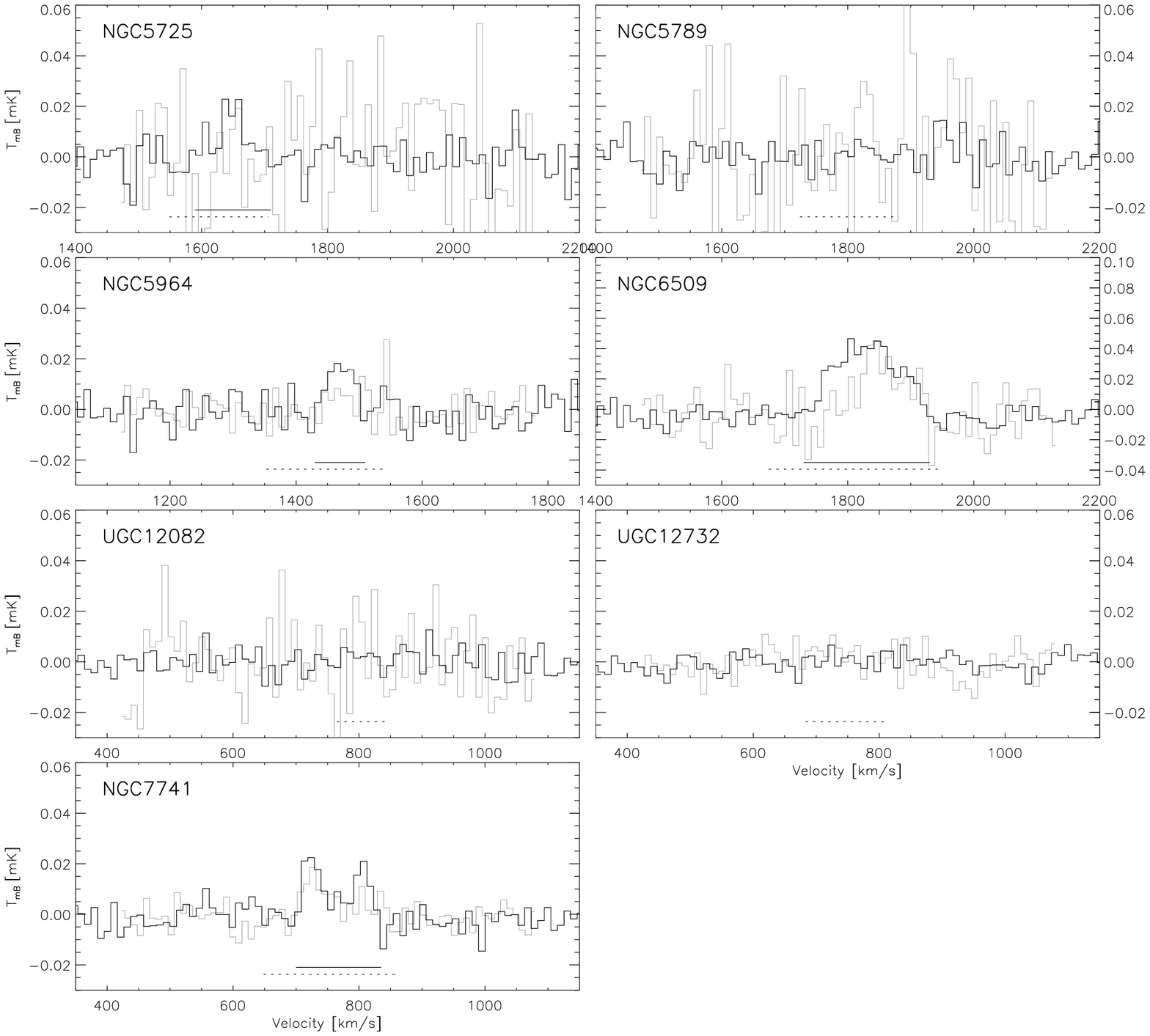}
\centerline{Fig.1 (cont.)}

\begin{figure}
\centering
\resizebox{0.5\hsize}{!}{\rotatebox{0}{\includegraphics{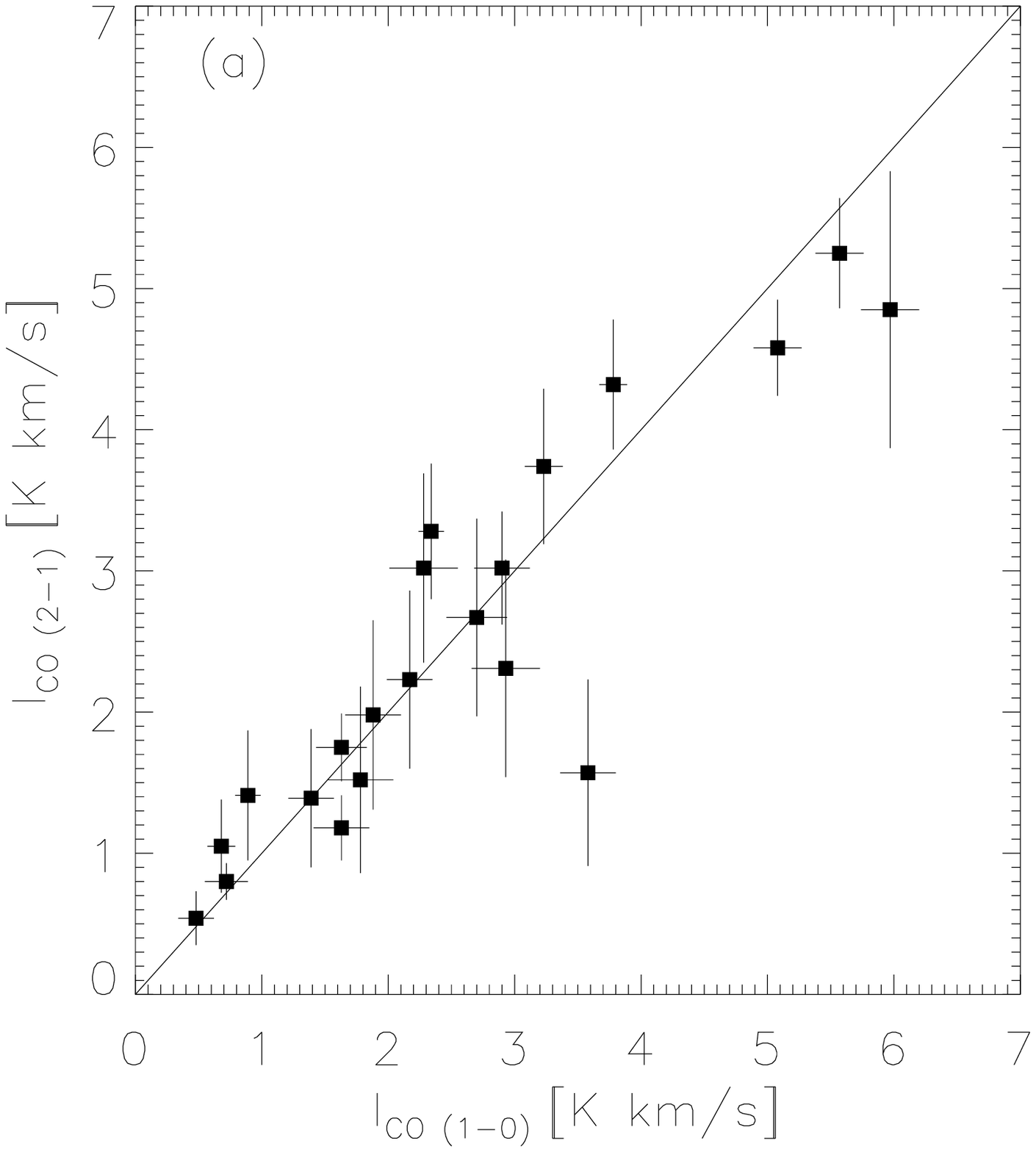}}} 

\vspace*{0.3cm}
\resizebox{0.54\hsize}{!}{\rotatebox{0}{\includegraphics{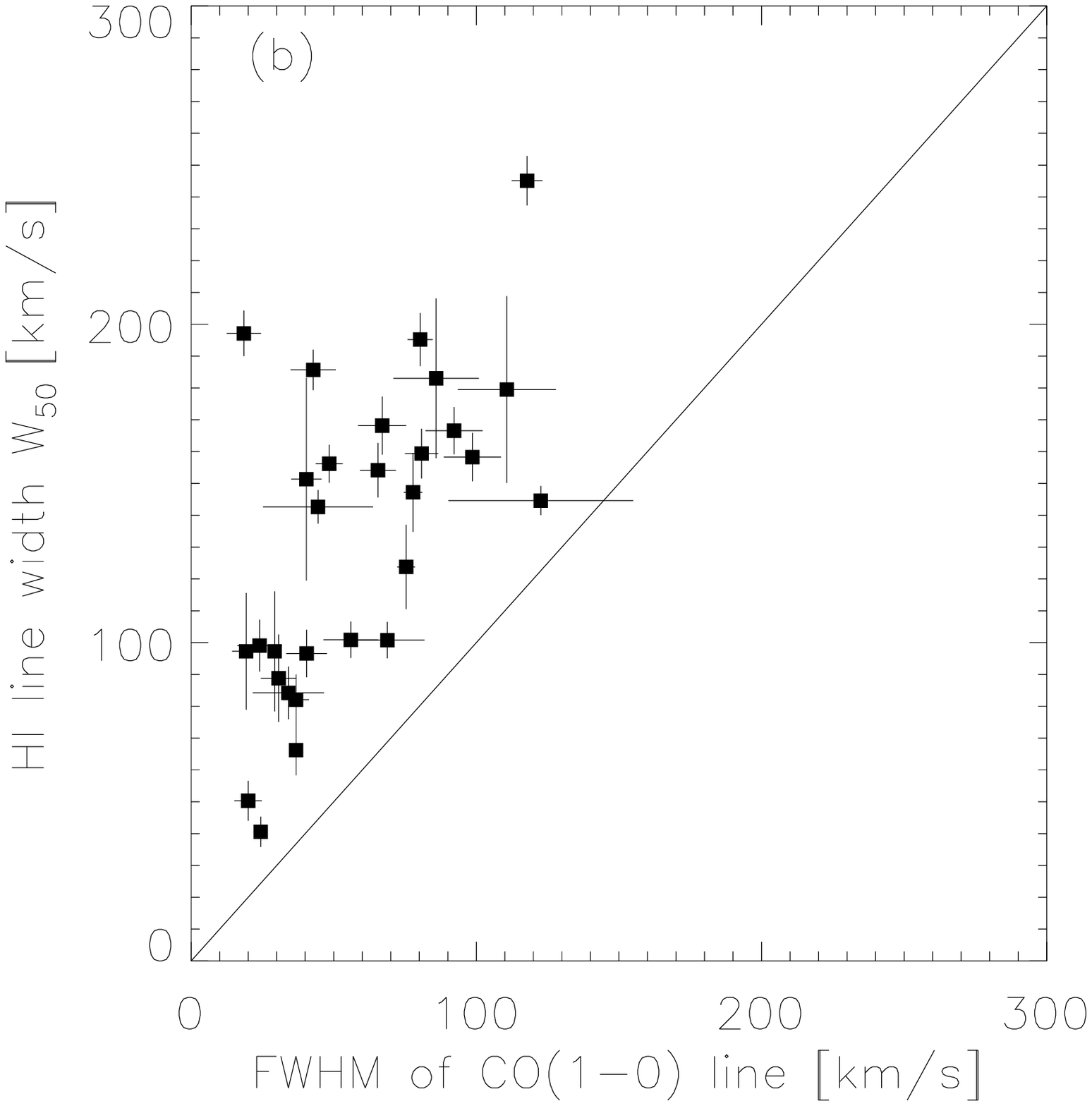}}} 
\caption{a) Comparison of observed line intensities for the \coone\ and \cotwo\
  transitions. The good agreement of both line intensities suggests that
  the CO gas extends over the area covered by the $3\mm$ IRAM beam ($21\as$). 
  b) FWHM of a Gaussian fit to the \one\ line 
  (Col.~4 of Table~\ref{tab:res}), compared to the HI $21\cm$ line 
  width, measured at 50\% of the peak intensity (from the LEDA database). 
}
\label{fig:lineratio}
\end{figure}

%
In Fig.~\ref{fig:lineratio}a, we compare the velocity-integrated 
intensities of the two CO lines. 
Since the telescope beam size is different for the two CO lines, 
their respective intensities are averaged over different areas 
in the center of the galaxy.
Under the assumption of optically thick molecular gas that extends
at least as far as the (larger) \one\ beam, one would expect 
an intensity ratio \ione /\itwo = $1$. If the CO emission 
were optically thick but more concentrated than the \coone\ beam,
one would expect \ione /\itwo $<1$  because beam dilution would lower
the observed \coone\ beam temperature.
The mean value for our sample is \ione /\itwo\ = $0.95 \pm 0.24$,
consistent with the assumption of optically thick, extended gas. 
This result is not atypical for spiral galaxies: \cite{bra93}  
found  \ione /\itwo\ = $0.89 \pm 0.06$ for 
a mixed sample of 81 galaxies. These authors mapped the
\cotwo\ emission in order to survey the same area as covered by
the \coone\ beam. Their result is therefore not affected by beam
dilution and demonstrates that the assumption of optically thick
gas is reasonable.

\begin{figure}
\centering
\includegraphics[width=0.5\hsize]{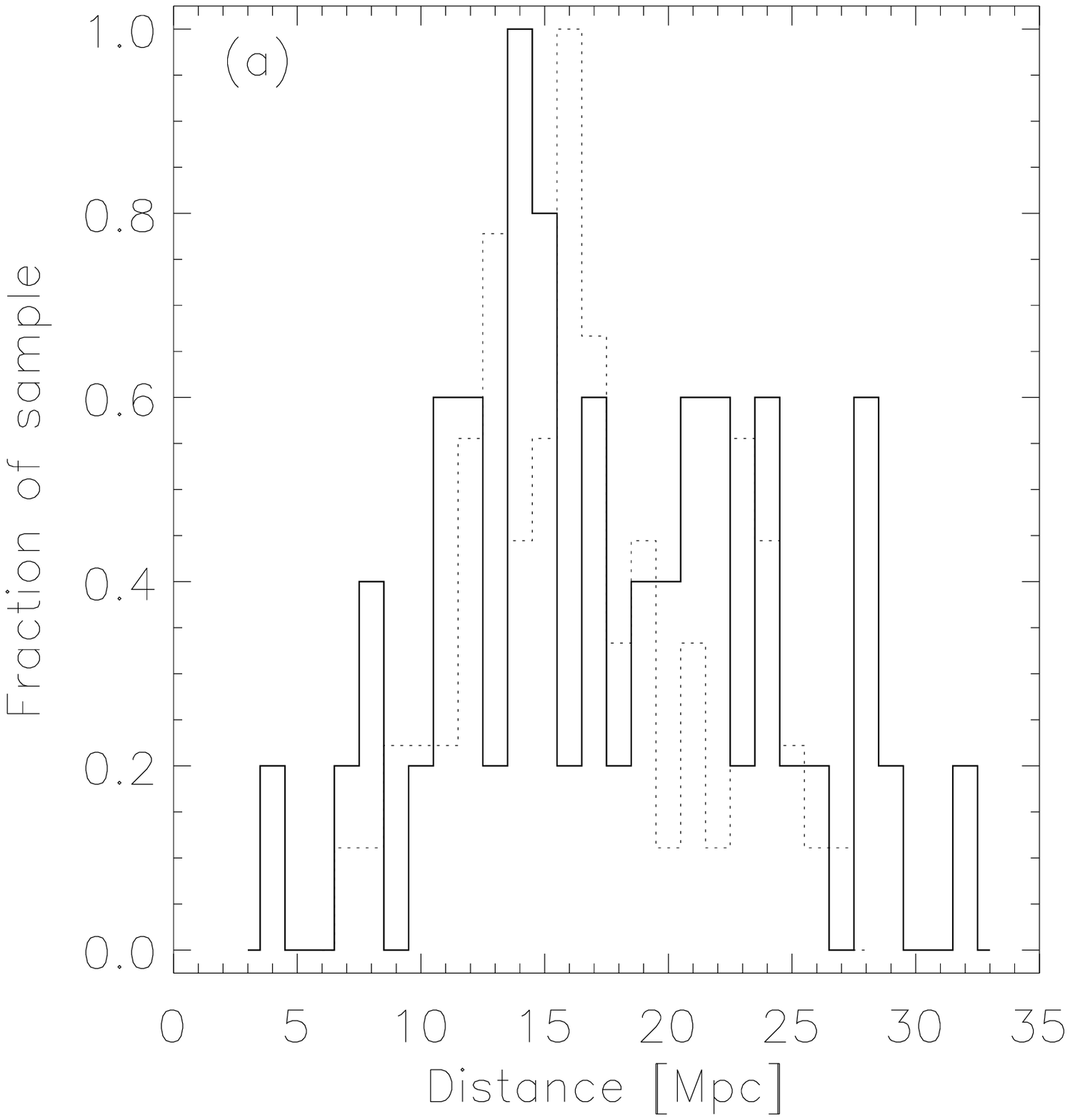} \\
\includegraphics[width=0.5\hsize,angle=-90]{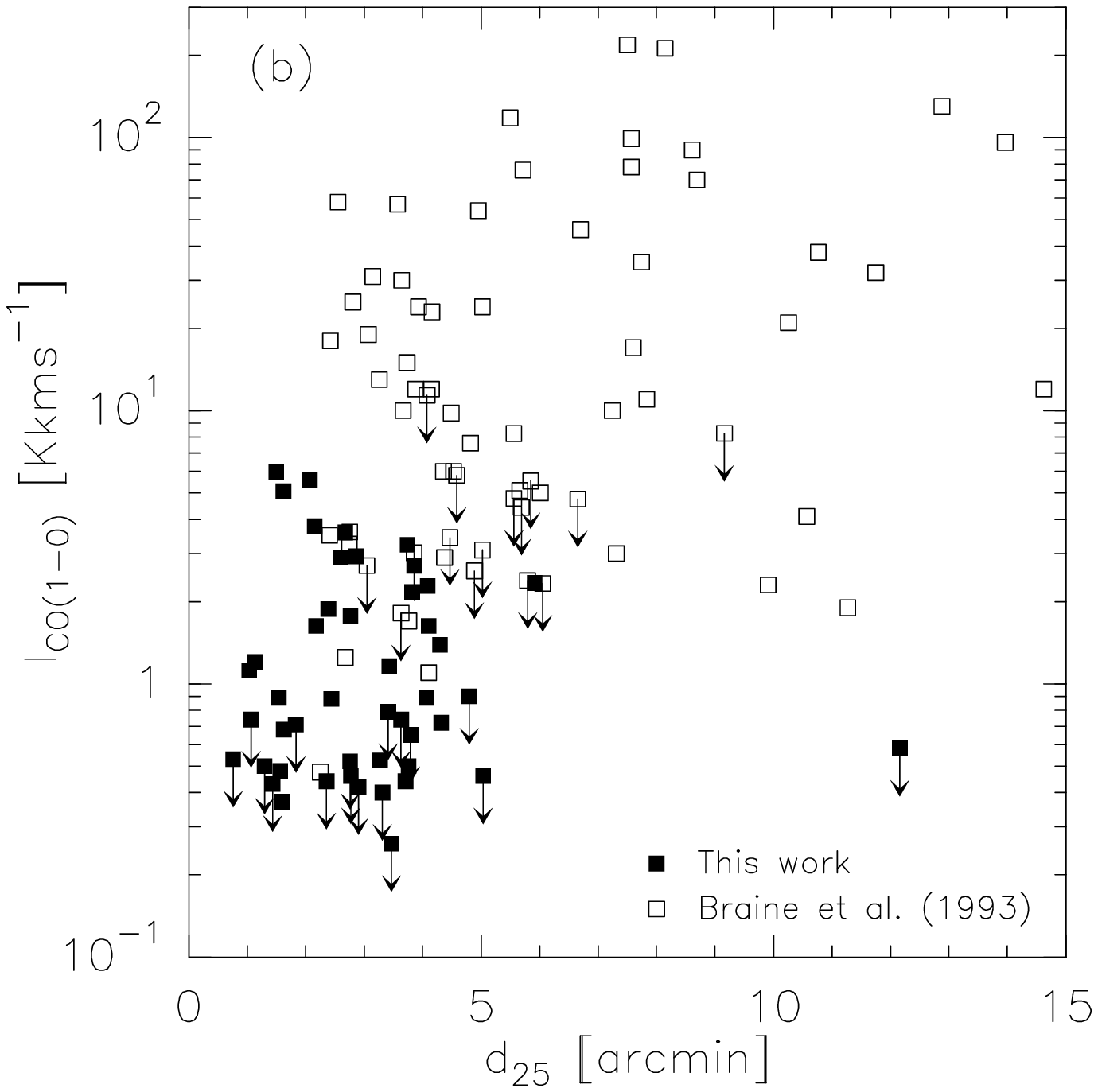}
\caption{
  a) Normalized histogram of galaxy distances for the \cite{bra93} 
  (dashed line) and our sample (solid line). The distance distributions 
  are very similar.
  b) Intensity of the \coone\ line as a function of angular diameter for 
  our sample (filled squares) and that of \cite{bra93} (open squares).
  See text for discussion. 
}
\label{fig:diam}
\end{figure}
For our galaxy sample, the CO line width is systematically smaller 
than the HI line width, as demonstrated in Fig.~\ref{fig:lineratio}b. 
This result differs from that obtained by \cite{sak99} who find that
the CO line width reaches about 95\% of the $W_{20}$ value within
$1\kpc$ from the center in a sample of 20 CO-luminous, 
mostly early- to intermediate-type spirals. However, given that
the bulge-less, disk-dominated systems in our sample have slowly 
rising rotation curves, and the beam size of our observations only covers 
$\approx 1.7\kpc$ at the median distance of our galaxy sample ($16.7\mpc$), 
this difference is not surprising. The smaller CO line width also
argues against the presence of strong non-circular motion
associated with prominent stellar bars in the inner $2\kpc$. 
%
\subsection{Molecular and atomic gas masses}\label{subsec:gasmass}
The amount of molecular hydrogen in external galaxies can not be
measured directly from \htwo\ emission, but has
to rely on indirect methods. The most common of these uses the \coone\
line emission as a tracer for \htwo , assuming proportionality between
the abundances of the two species. Driven mostly by the lack of a clear 
theoretical understanding of the relative abundances, many studies have 
assumed that the conversion factor $X\equiv N({\rm H_2})/I({\rm CO})$ is 
universal, i.e. applies to all environments. 

Keeping in mind the shortcomings of this approach, we can 
estimate\footnote{Here, we have
adopted a Gaussian beam for which the solid angle is $1.13 \cdot \Theta^2$. 
For comparison, a circular beam (which is also used in the literature) 
covers a solid angle of $\pi /4 \cdot \Theta^2$, and hence yields 
$M_{\rm H_2} = 68 \cdot D^2\cdot I_{10}\cdot \Theta^2  [\msun]$. }
the \htwo\ mass within the central $21\as$ by 
applying a generic conversion factor of 
$X = 2.3 \cdot 10^{20} \cm^{-2}(\kkk)^{-1}$ \citep{str88} which yields:
\beq \label{eq:mh2}
M_{\rm H_2} = 97 \cdot D^2\cdot I_{10}\cdot \Theta^2 \>\>\>\> [\msun]
\eeq
where the galaxy distance $D$ is measured in Mpc, $\Theta$, the half power 
beam width of the 30\,m-telescope ($21\as$) in seconds of arc,  
and \ione \ the velocity integrated \coone\ intensity in $\kkms$.
The distance estimates for our sample are obtained by using 
the measured recession velocities, corrected for Virgo-centric 
infall \citep{san90}, and assuming $\rm H_0 = 70\kms Mpc^{-1}$.
In the case of NGC~5584, for which only the \two\ line was detected, 
we estimate \ione\ by assuming that the \one\ measurement is affected by
beam dilution, i.e. the CO emitting region is small compared to the \two\ 
beam. In this case, if the intrinsic \ione/\itwo\ ratio is unity,
as appears to be the case for the rest of the sample, then \ione=\itwo/4
in NGC\,5584. This value is consistent with the upper limit derived from 
the \one\ spectrum itself.

The resulting molecular gas masses for the detected galaxies
(Col.~6 of Table~\ref{tab:res}) 
span the range between $1.2 \times 10^6$ and $1.9 \times 10^8$ $\msun$.
We emphasize that these mass estimates represent only lower limits 
to the total molecular gas content, since the galaxies were only
observed at their center position (on average, the IRAM 
beam diameter is only about 10\% of $d_{25}$).

There is general agreement, both from models \citep{mal88,kau99} and from
observations  that $X$ is not constant but in fact depends on environmental 
parameters such as metallicity, density, or the radiation environment.
Unfortunately, observations have not yet yielded a unique 
prescription  for these dependencies. Studies of nearby
galaxies provide evidence for a strong dependence of $X$ on 
metallicity, but the various authors report different slopes 
\citep{wil95,ari96,isr97,bos02}. 

In order to test the robustness of our analysis resulting from
the use of Eq.~(\ref{eq:mh2}), we also apply a galaxy-dependent $X$. 
Specifically, we use the relation suggested by \cite{bos02} which was 
derived from a sample of 14 well-studied nearby galaxies. Their relation
between $X$ and the metallicity agrees well with the result of
\cite{ari96}, with a slope that is stronger than the one found by 
\cite{wil95}, and weaker than that derived by \cite{isr97}. 
For our purposes, we use the relation between $X$ and the blue
magnitude of \cite{bos02}: 
\beq \label{eq:bos}
{\rm log}(X) = (0.18\pm 0.04) \cdot {\rm M_B} + (23.77\pm 0.28)
\eeq
The resulting \htwo\ masses, \mhtwo(var) 
are also listed in Table~\ref{tab:res} 
(Col.~7).

In order to compare the mass of the molecular gas component to 
that of the atomic one, we calculate the total HI mass for our 
galaxy sample from the 21\,cm flux (as listed in LEDA) following 
the approach described in \cite{dev91}, Eq. (74-77). The HI masses 
derived in this fashion represent the total HI mass of the galaxy,
they are listed in Col.~8 of Table~\ref{tab:res}. 
%
\section{Analysis} \label{sec:analysis}
\subsection{Comparison to early-type galaxies} \label{subsec:compare}
One of the goals of this program is
to complement existing studies of the molecular gas content of
spiral galaxies with a sample of ``pure'' disk galaxies with Hubble
types around Sd. These galaxies are underrepresented in existing
samples, and it is important to adress the question whether they 
follow the same trends as intermediate- to early-type spirals,
or whether they might have intrinsically different gas properties. 
We have therefore compared our results to those obtained by 
\cite{bra93} for a sample of 81 predominantly early-type spiral galaxies. 
Their dataset is particularly suited for comparison with ours
because both samples have nearly identical distance distributions
(Fig.~\ref{fig:diam}a) and were observed with the IRAM $30\m$ telescope.
In particular, the \coone\ intensities of the \cite{bra93} sample 
also refer to the central position of the galaxies only, and are
therefore directly comparable to our data.
The \cite{bra93} sample was selected 
to include nearby spiral galaxies with $\rm m_B < 12\mags$.
In contrast, our sample mostly contains fainter galaxies
(42 out of 47 galaxies have $\rm m_B > 12\mags$).
Fig.~\ref{fig:diam}b shows that the angular diameter of the major 
axis, $d_{25}$, of the \cite{bra93} galaxies is on average larger 
than for our targets, although both samples have nearly identical distance 
distributions. This is to be expected, because late-type spirals have on 
average lower total luminosities and lower surface brightness \citep{rob94}, 
both of which likely reduce the isophotal diameter. 

As evident from Fig.~\ref{fig:diam}b, neither sample shows an 
obvious trend between the angular size of the galaxy disk and \ione .
We have also verified that there is no correlation between
the angular size and  \mhtwo\ or \mhtwo(var). 
This indicates that the CO observations are not significantly biased by the
ratio of the optical disk size to the IRAM beam.
Since the optical diameters in both samples are much larger than  
the IRAM \one\ beam of $21\as$, it is likely that in both studies,
the observations miss a non-negligible fraction of the total CO emission.
This is confirmed for those 6 galaxies in our sample for which wide-beam 
data are listed in the GOLDMine database \citep{gav03}. For these 6 objects, 
the IRAM beam detects only between 15 and 60\% of the total \ione . 
In principle, the undetected CO fraction should increase with 
galaxy diameter because the size of the CO-emitting region and the 
diameter of the optical disk are correlated 
\citep[$d_{\rm CO}/d_{25} \approx 0.5$][]{you95}.
However, the fact that the central \ione\ is on average
larger for the \cite{bra93} sample demonstrates that typically,
early type spirals have more CO within their central few kpc than 
late type spirals.

\begin{figure}
\centering
\includegraphics[width=0.5\hsize,angle=-90]{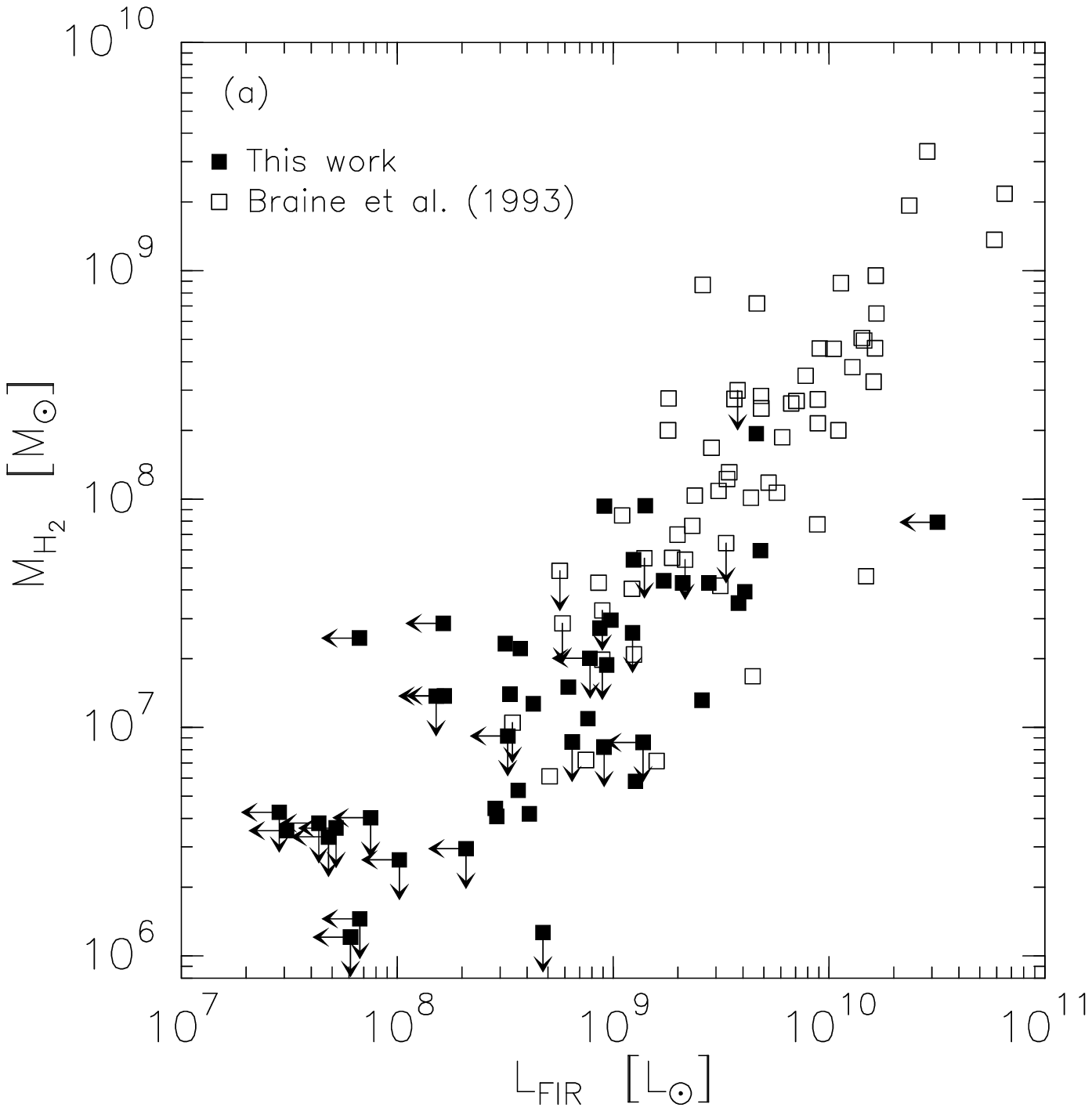} 

\vspace*{0.4cm}
\includegraphics[width=0.5\hsize,angle=-90]{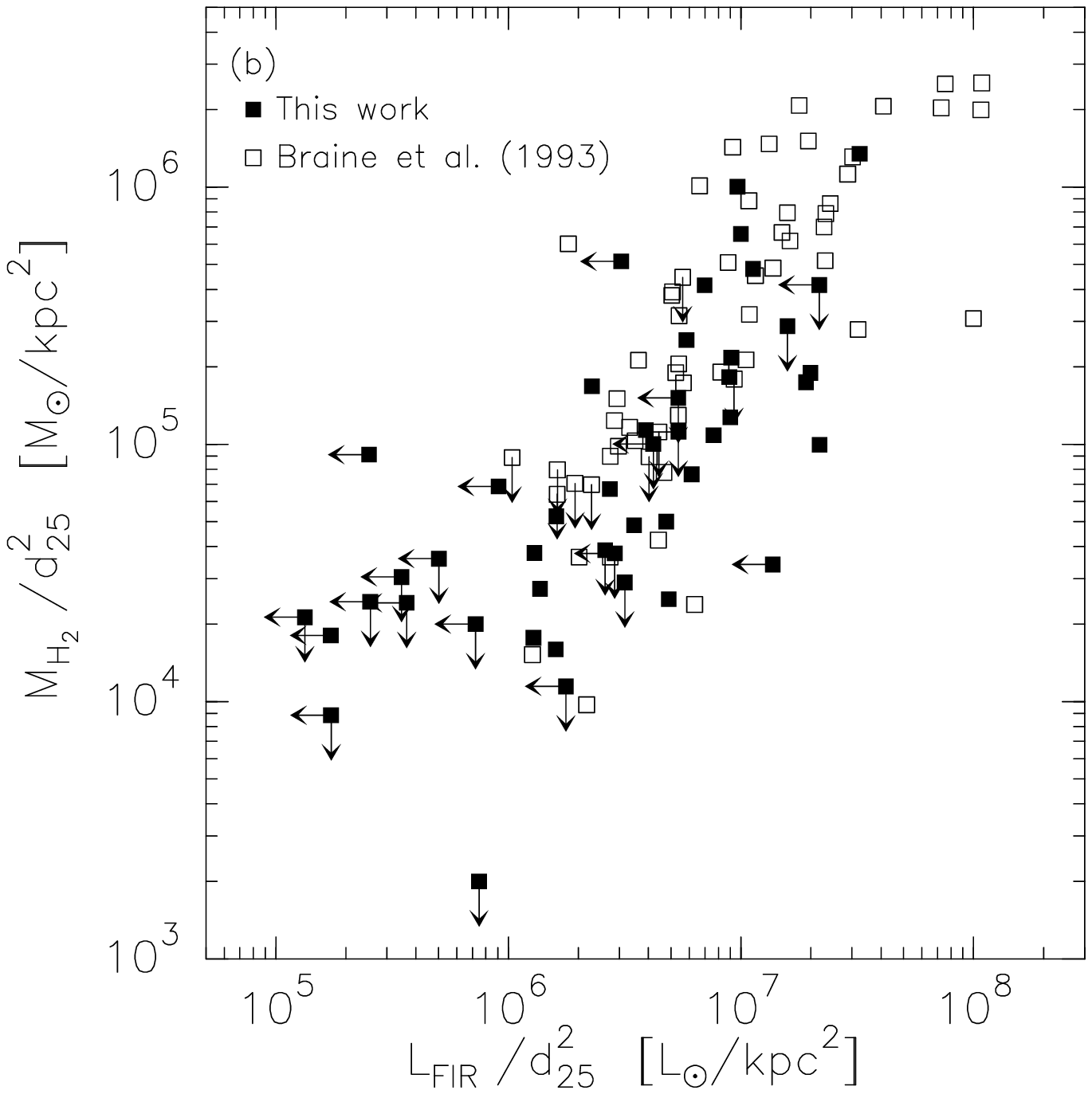}
\caption{  
  a) \htwo\ mass as a function of far-infrared luminosity (taken from LEDA)
  for our galaxy sample (filled squares), compared to that of \cite{bra93} 
  (open squares). The latest-type spirals follow the same relation established 
  for intermediate- and early-type spirals, and extend it to lower 
  luminosities.
  For galaxies in our sample without IRAS measurements in LEDA, we have used 
  the SCANPI tool provided by the NASA/IPAC Infrared Science Archive (IRSA) 
  to derive an upper limit to the FIR luminosity.
  b) same as a), but \htwo\ mass and far-infrared luminosity normalized 
    to the galaxy surface area.
}
\label{fig:fir}
\end{figure}

In Fig.~\ref{fig:fir}a, we add the results of our survey
to the well-known relation between molecular gas mass (here calculated 
with the generic value for $X$, but using Eq.~(\ref{eq:bos}) 
yields equivalent results) and the far-infrared (FIR) luminosity
\citep[e.g.][]{you91}. The galaxies of our sample closely follow
the relation defined by the earlier-type spirals, and extend it towards 
lower FIR-luminosities. 
The slope of the correlation is consistent with 1, indicating
direct proportionality between the two quantitites. The average
value of \lfir/\mhtwo\ is about 40. This is higher than in
nearby normal spirals (\lfir/\mhtwo $\simeq 5$) and even starburst
galaxies (\lfir/\mhtwo $\simeq 15$) \citep{san96}, another indication
that measurements with a single IRAM pointing miss a substantial
fraction of total molecular gas.
It is interesting that after normalization of the two quantities 
to the optical diameter of the galaxy, the two samples overlap and
cover a similar range (Fig.~\ref{fig:fir}b).
This suggests that the latest-type spirals follow the same physical 
processes that intimately link the molecular gas with star formation activity. 
At least in this context,
the latest-type spirals are not a distinct class of objects, but rather 
constitute the low-luminosity end of the galaxy distribution.
%

\begin{figure*}
\includegraphics[width=\hsize]{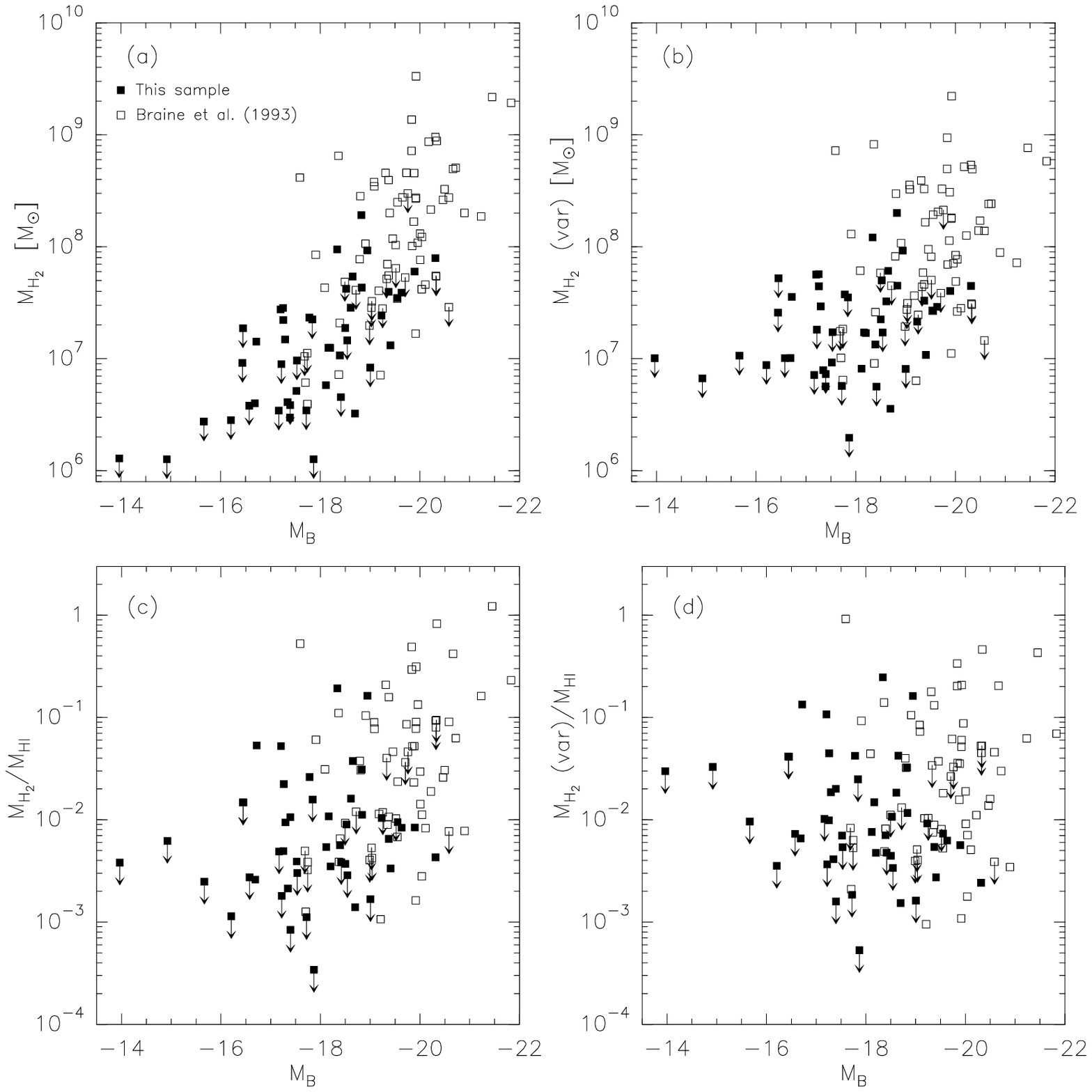}
\caption{
  \htwo\ mass (top) and \mhtwo/\mhi\ ratio (bottom)
  as a function of $\rm M_B$ for our galaxy sample (filled
  squares), compared to that of \cite{bra93} (open squares).
  Here, the \htwo\ mass has been calculated using both the
  generic Galactic conversion factor (Eq.~(\ref{eq:mh2}), left) 
  as well as the luminosity-dependent $X$ of \cite{bos02} 
  (Eq.~(\ref{eq:bos}), right).
}
\label{fig:mb}
\end{figure*}

\begin{figure*}
\includegraphics[width=\hsize]{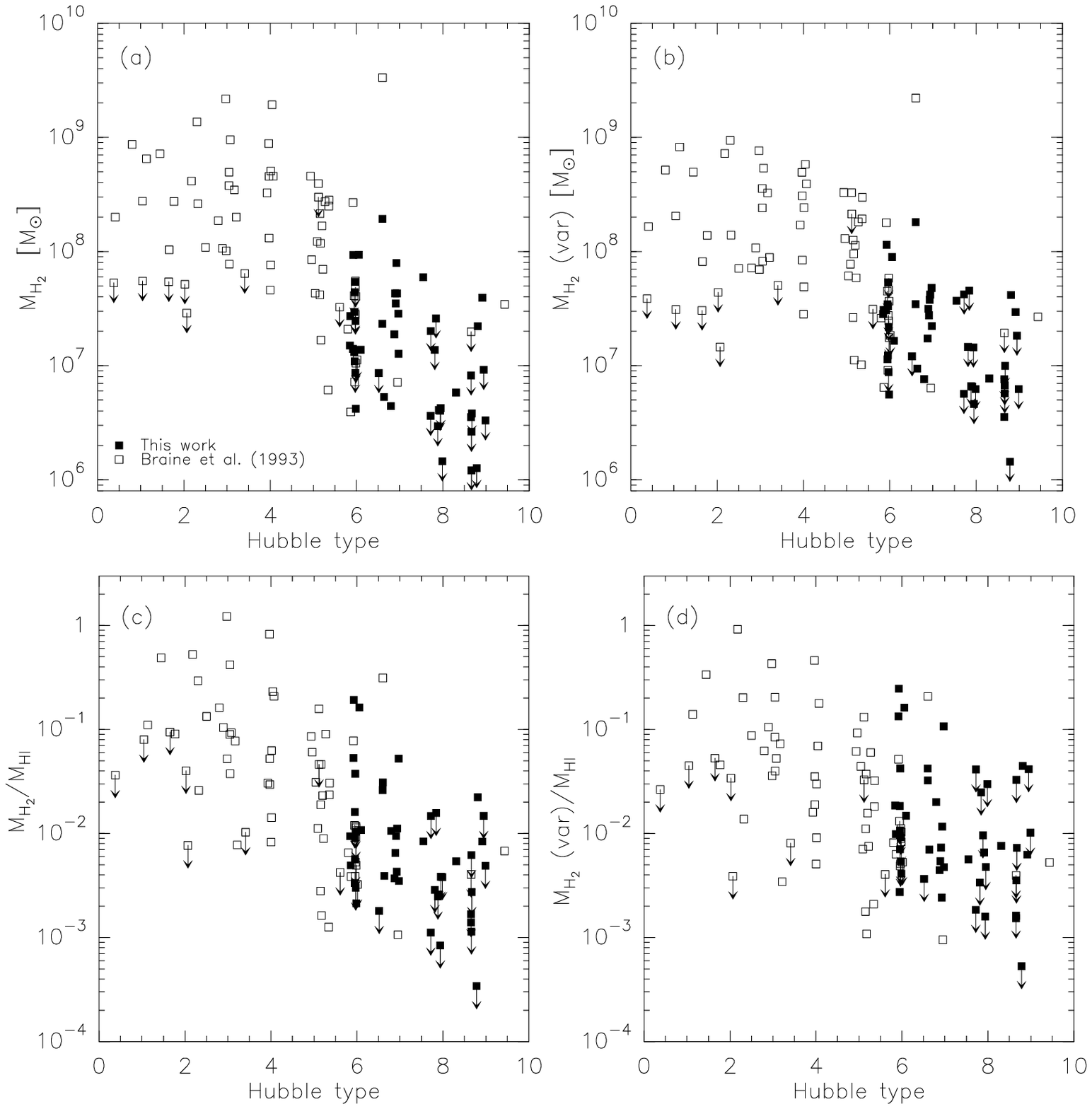}
\caption{
  \htwo\ mass (top) and \mhtwo/\mhi\ ratio (bottom)
  as a function of Hubble-type (as listed in LEDA)
  for our galaxy sample (filled squares), compared to that of 
  \cite{bra93} (open squares).
  As in Fig.~\ref{fig:mb}, we calculate the \htwo\ mass  
  using the Galactic value for $X$ (left) as well as
  the luminosity-dependent $X$ of \cite{bos02} (right).
}
\label{fig:hubbletype}
\end{figure*}

A similar conclusion can be reached from the
relation between \htwo\ mass and total magnitude \mb\ shown in
Fig.~\ref{fig:mb}: optically fainter galaxies have less molecular gas
in their central regions. 
This result holds also when calculating the molecular mass according
to Eq.~(\ref{eq:bos}), although this increases the
scatter and reduces the dynamical range of \mhtwo .
A similar, albeit weaker, correlation 
exists between optical luminosity and molecular-to-atomic gas mass ratio. 
Using Eq.~(\ref{eq:bos}) for $X$ again 
increases the scatter, so that only a weak trend remains  
(Fig.~\ref{fig:mb}c, d). Here, as well as in Fig.~\ref{fig:hubbletype}c
and d, we have excluded three galaxies (NGC\,2681, NGC\,4274, and NGC\,4438)
with anomalously\footnote{Specifically, we consider as
HI-deficient those galaxies which deviate by more than $3\sigma$ from
the relation between \mhi\ and linear diameter found by \cite{hay84} (their
Table\,V).} low HI content from the \cite{bra93} sample.

We find that the latest-type spirals have on average less molecular 
gas and lower molecular gas mass fractions in their central regions than 
their early-type cousins. This is demonstrated in Fig.~\ref{fig:hubbletype} 
which plots the \htwo\ mass and \mhtwo/\mhi\ ratio versus Hubble-type. 
This result seems robust against variations in $X$, although 
using a luminosity-dependent conversion factor increases the scatter and 
decreases the slope of the correlations (Fig.~\ref{fig:hubbletype}b, 
d). The dependency of molecular gas mass fraction on Hubble-type
and luminosity apparently contradicts the results of \cite{bos02}
who find that the fraction of molecular gas is independent of Hubble type or 
luminosity. It should be kept in mind that for our survey - as well as
for the \cite{bra93} data - the IRAM beam is generally much smaller than 
the optical or HI disk of the galaxy. This probably explains the fact that 
typical gas mass fractions in Fig.~\ref{fig:hubbletype}c)
and d) are only a few percent, much smaller than the $\sim$15\% found by 
\cite{bos02}. However, effects related to beam size cannot explain the 
reduced \htwo\ fraction at the late end of the Hubble sequence in 
Fig.~\ref{fig:hubbletype}c) and d) because both samples were observed 
in the same way with the same instrument and have
similar distance distributions (Fig.~\ref{fig:diam}a).
Furthermore, Fig.~\ref{fig:diam}b shows that the optical diameter $d_{25}$ 
is on average smaller for late-type spirals than for earlier Hubble types, and
the $21\as$ IRAM beam therefore covers a larger fraction of the optical disk.
If the extent of molecular gas scales with the optical diameter
in a way that is common to all galaxies, one would expect our sample
to have a {\it higher} molecular gas fraction than the \cite{bra93} sample.

The fact that we observe the opposite result therefore indicates that the 
molecular gas in late-type spirals is less centrally concentrated relative 
to the HI distribution than in early-type spirals if indeed the overall
\htwo /HI ratio is the same for all types of spirals, as indicated by
the results of \citet{bos02}.
This is in qualitative agreement with the result of \cite{you95} 
who find that the ratio between the CO and optical isophotal radius
increases with Hubble type. 

As discussed by \cite{boe03}, one should not place too 
much emphasis on the exact Hubble classification in the range Scd-Sm 
because at arcsecond resolution, the prominent nuclear star cluster is 
easily mistaken as a compact ``bulge'', and hence any morphological 
classification based on ground-based images is inaccurate at best. 
Nevertheless, it is clear from Fig.~\ref{fig:hubbletype}a) and b) that
late-type galaxies are rather inefficient in accumulating molecular
gas in their central regions.
Similar results have been found by a number of earlier studies 
\citep[e.g.][]{you89,sag93,cas98}. \cite{you89} have suggested 
that the deep gravitational well of a prominent bulge facilitates 
the formation and/or accumulation of molecular gas. While this might
well be true, it is certainly not the whole story. Our sample
has been selected to be devoid of stellar bulges, yet it shows a wide
range of \htwo\ masses. In the next section, we use the high-resolution
HST images to investigate the impact of the shape of the stellar disk
on the amount of molecular gas that accumulates in the center.

The results presented in this section depend to some degree on the 
assumptions made for $X$, in particular the variation with
metallicity and/or luminosity which is still a matter of debate.
For example, a much stronger dependence of $X$ on metallicity than 
the one derived by \cite{bos02} has been suggested by \citet{isr97}. 
Using Fig.~2 of \citet{bos02}, we can infer that the variation in $X$
across the range of \mb\ covered by the galaxies discussed here
(-14 $\geq$ \mb\ $\geq$ -21) is equivalent to that caused by
metallicity differences of about 1 dex. According to \citet{isr97},
$X$ would vary by a factor of 500 over this metallicity range. 
While such a high variance in $X$ seems somewhat unlikely, it 
would eliminate (or even reverse) the trends of \mhtwo\ or \mhtwo/\mhi\ 
with \mb\ and Hubble type in Figs.~\ref{fig:mb} and \ref{fig:hubbletype}.
%
\subsection{CO and the stellar disk}\label{subsec:disk}
\begin{figure}
\centering
\resizebox{0.5\hsize}{!}{\rotatebox{270}{\includegraphics{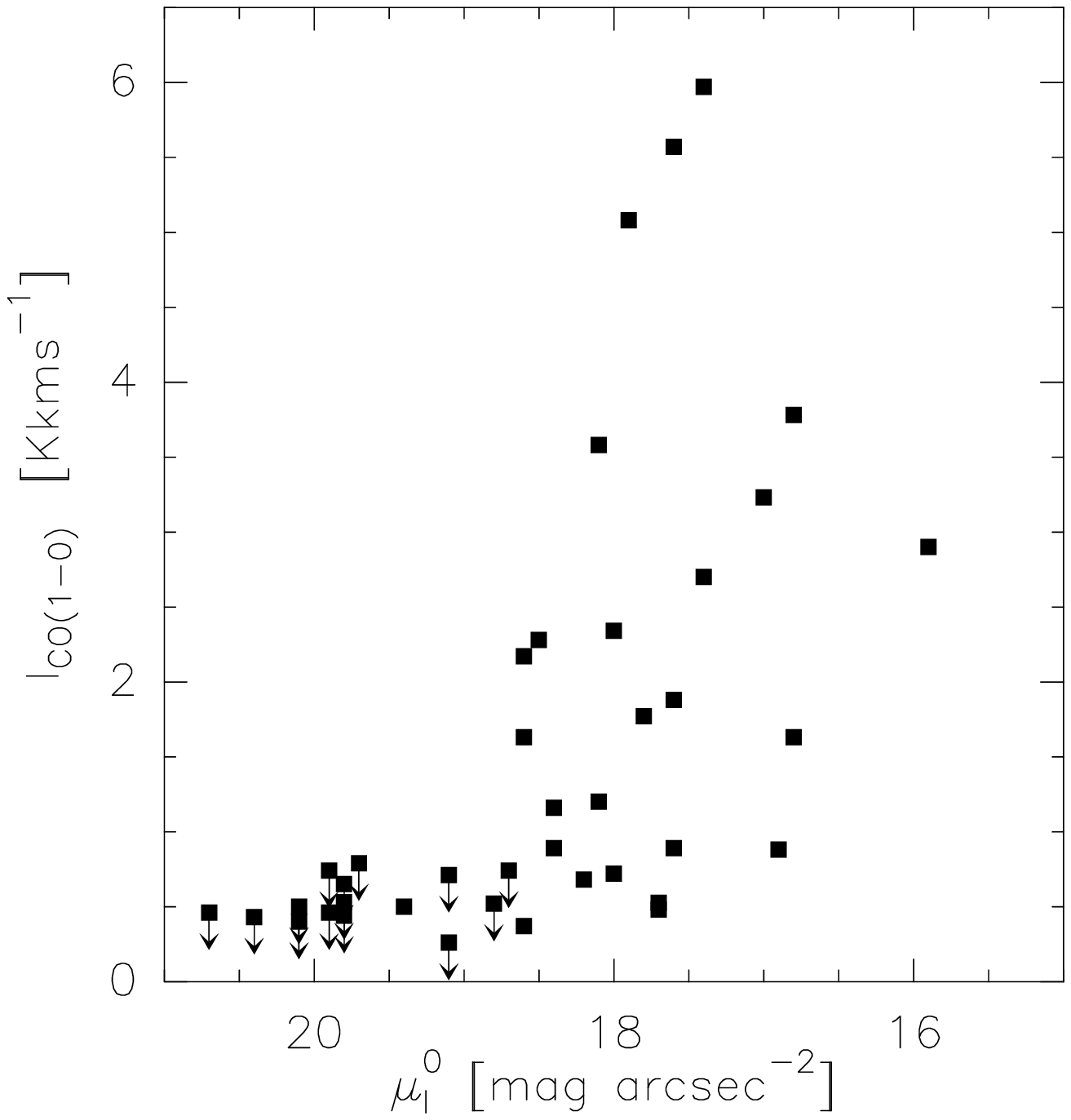}}} 
\caption{  
  The \coone\ line intensity as a function of $\mu_I^0$, the central 
  I-band surface brightness of the galaxy disk (i.e. underneath the 
  nuclear star cluster). Nearly all galaxies detected in CO have 
  $\mu_I^0 < 19\mags$.
}
\label{fig:sb0_vs_ico}
\end{figure}

In order to measure the central surface brightness of the galaxy disk
{\it underlying the \nc } we have used the results of the elliptical
isophote fits presented in paper~I. Here, we have adopted the average
of the two inward extrapolations of the disk shown in Fig.\,3 of
paper~I. In Fig.~\ref{fig:sb0_vs_ico}, we compare the so-derived
central disk surface brightness $\rm \mu_I^0$ which is directly
proportional to the average molecular gas surface density \ione\
within the beam.
Both quantities are independent of galaxy distance, as long as the CO is
more extended than the IRAM beam, which - based on the observed
CO line ratio discussed in Sect.~\ref{sec:results} - is most likely 
true for the majority of our sample.

The most direct result of Fig.~\ref{fig:sb0_vs_ico} is that there
appears to be a threshold surface brightness for detection at the
sensitivity limit of our observations:
we detect all 25 galaxies with $\mu_I^0 < 18.7\mpas$, 
but only 1 out of 15 galaxies with $\mu_I^0 > 18.7\mpas$.
This trend cannot plausibly be explained by anomalously 
high values of $X$ in the undetected galaxies, because their absolute
magnitudes span a wide range (-14 $\geq$ \mb\ $\geq$ -19.5).
Figure~\ref{fig:sb0_vs_ico} therefore indicates that the stellar density 
in the central regions 
of late-type spirals is intimately linked to the molecular gas abundance,
even in the absence of a massive stellar bulge. It is difficult to
assess whether there is a direct correlation between $\rm \mu_I^0$ and \ione .
To address this issue, deeper CO observations for a larger galaxy 
sample which expands the range of surface brightness levels would be 
required. Nevertheless, this result might help to explain the low success 
rate in detecting molecular gas in low surface brightness (LSB) galaxies.
This class of galaxies, defined to have $\rm \mu_B^0 > 23\mpas$
or $\rm \mu_I^0 > 21-22\mpas$ for typical disk colors \citep{dej96},
has only been successfully detected in 3 out of 34 attempts 
\citep[see compilation by][]{one03}.
The typical CO intensity of the detected LSB galaxies ($\sim 1.2\kkms$)
is similar to that of our sample (both datasets were obtained with
the $30\m$ telescope, and are thus directly comparable). 
However, the upper limits 
reported by \cite{one03} for non-detections ($< 0.3\kkms$) fall
well below our sensitivity limit. It is therefore possible that
LSB galaxies would extend the correlation between central surface
brightness and CO intensity to fainter levels.

\subsection{CO and nuclear star clusters}\label{subsec:cluster}
Our survey has shown that the latest-type spirals are not entirely
devoid of molecular gas. With a median \htwo\ mass of 
$1.4\times 10^7\msun$ inside the IRAM beam, there is 
enough raw 
material within the central kpc to support a number of modest 
($\rm 10^5\,\msun$ in stars) starburst episodes assuming a
star formation efficiency of 10\%. 
This is an important
result because recent spectroscopy of nuclear clusters has shown that
the spectral energy distribution (SED) of a large fraction of these objects
is dominated by a relatively young population of stars with an age of
a few hundred Myrs \citep{wal03}. Because it is unlikely that we
witness the first nuclear starburst in such a large number of
galaxies, it is reasonable to assume that star formation episodes
within the central few pc occur repetitively. If true, there must be a
reservoir of molecular gas to support these events. While it is
comforting that our observations have demonstrated the presence of
reasonable amounts of molecular gas in the central regions of late-type
spirals, much has to be learned about the details of nuclear cluster 
formation. Recent high
resolution ($\sim 10\pc$) CO observations of the barred late-type spiral
IC\,342 demonstrate that gas can indeed accumulate inside the central
few pc \citep{sch03}. In the case of IC\,342, rough estimates of
the gas inflow rates suggest that repetitive nuclear starbursts can
be supported with duty cycles between $\sim 100\myr$ and $1\gyr$. 
However, IC\,342 presents only a case study, and it is uncertain
whether similar processes are common in late-type galaxies. 

\begin{figure}
\centering
\resizebox{0.5\hsize}{!}{\rotatebox{270}{\includegraphics{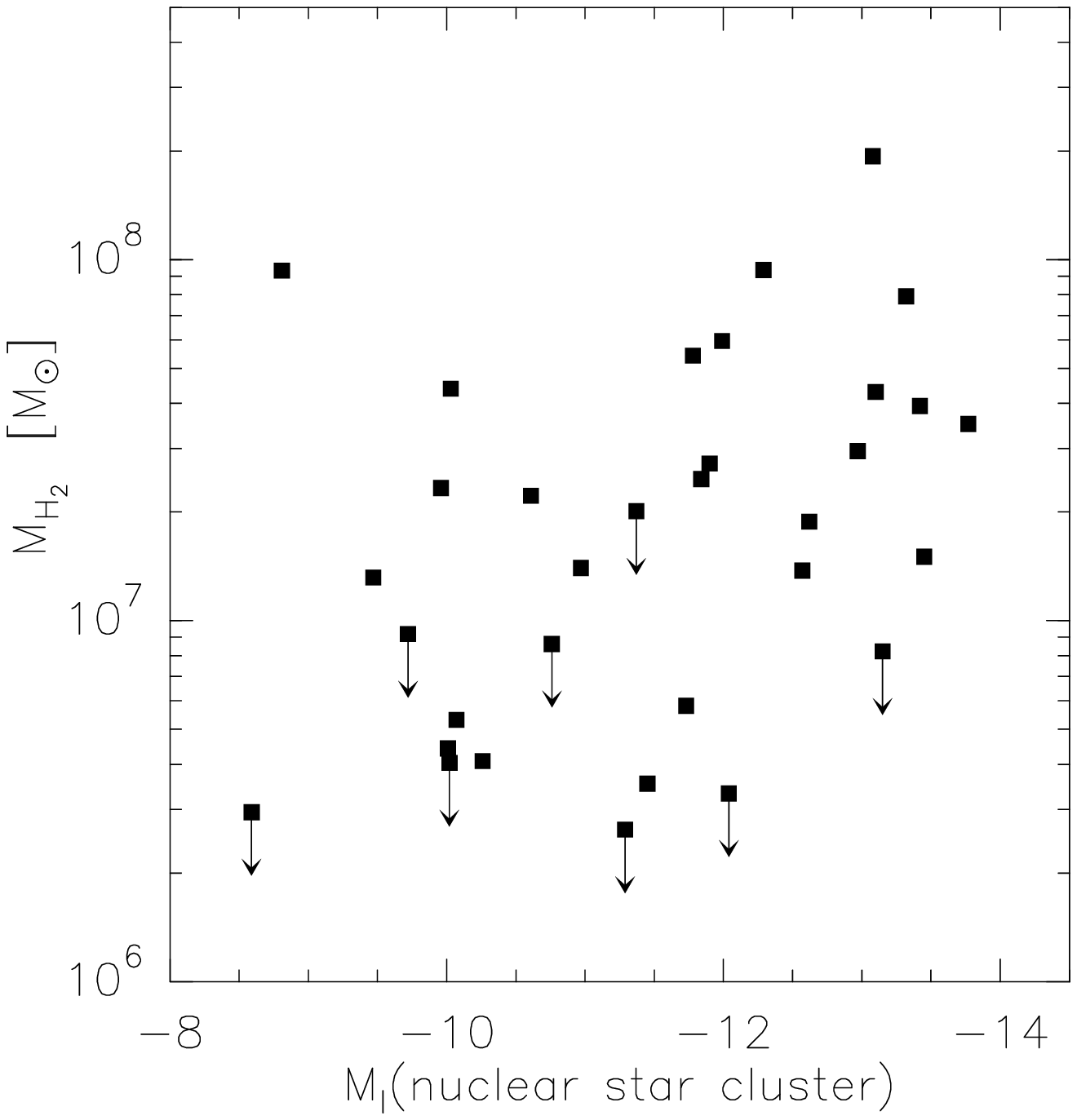}}} 
\caption{  
  \htwo\ mass as a function of the absolute I-band magnitude of the 
  nuclear star cluster. The CO detection rate appears higher for 
  galaxies with more luminous nuclear clusters.
}
\label{fig:clustermag_vs_ico}
\end{figure}

As we have discussed in the last section, the stellar density over
the central kiloparsec appears tightly linked to the molecular gas
abundance (or vice versa). On the other hand, the luminosity of the
nuclear star cluster appears largely independent of $\rm M_{H_2}$,
as demonstrated in Fig.~\ref{fig:clustermag_vs_ico}.
Again, this result is independent on the exact recipe to calculate
the \htwo\ mass: using Eq.~(\ref{eq:bos}) for the $X$ does not
significantly change Fig.~\ref{fig:clustermag_vs_ico}.
This result is not too surprising because the cluster luminosity is 
mostly determined by the age of the youngest population, and hence 
is not a reliable indicator of cluster mass (which one might expect to
be more tightly correlated with CO luminosity). 

However, it is interesting to note that $\sim$ 90\% of the galaxies
with nuclear clusters of $\rm M_I < -11.5$ are detected in CO, whereas
the detection rate is about a factor of three lower for less
luminous clusters. In addition, out of the 7 galaxies {\it
without} a nuclear star cluster only one (NGC\,7741) is detected in CO.
This result underlines the notion that the lack of molecular gas is
the main reason for the uneventful star formation history in the
central regions of these galaxies. A more thorough study of the connection 
between molecular gas supply and nuclear starbursts requires reliable mass 
and age estimates for the nuclear clusters. Such measurements - which
are now possible from high-resolution optical spectroscopy with large ground-based
telescopes \citep{wal03} - will undoubtedly provide new insights into the
formation mechanism of nuclear clusters. In the meantime, it is not 
unreasonable to speculate that the frequency and efficiency of nuclear 
starbursts is governed by the supply of molecular gas which in turn is 
regulated, at least in part, by the gravitational potential of the host 
galaxy disk.  
%
\subsection{Are bars important?} \label{subsec:bars}
As discussed so far, the gravitational potential of the stellar mass
distribution appears to play an important role for the gas flow
towards the nucleus. In this paragraph, we discuss whether our
observations provide evidence that dynamical effects such as gas flow
in a non-axisymmetric potential are also important. A way to gauge the
impact of stellar bars on (circum)nuclear star formation is to measure
the amount of molecular gas in the central kiloparsec in barred and
unbarred galaxies. A number of studies have shown that the nuclear gas
concentration in barred galaxies is indeed higher than in galaxies
without bars \citep[e.g.][]{sak99,she01,she03}, thus confirming the theoretical 
picture that bars drive molecular gas toward the galaxy center 
\citep[e.g.][]{ath92}. However, the (few) galaxies with late Hubble types 
in these studies were selected to be luminous in the blue or the FIR, i.e. 
they are biased towards high star formation rates. 

Our sample, on the other hand, has been selected only for Hubble type,
inclination, and distance (see Sect.~\ref{subsec:sample}). It should therefore
provide an unbiased look at the impact of stellar bars on the molecular gas 
abundance in the latest Hubble-type spirals. Our sample is evenly divided in 
barred, mixed, and unbarred galaxies following the RC3 \citep{dev91} 
classification scheme (morphological type SB, SAB, or SA, respectively).
In Table~\ref{tab:barstuff}, we
summarize the CO statistics for the three subsamples. We detect 80\%
of the barred galaxies in our sample, compared to 63\% of mixed and
44\% of unbarred galaxies. Taken at face value, this result supports 
the notion that bars are indeed an important factor for the transport of 
molecular gas towards the central few kiloparsec, even in very late 
Hubble types. However, we caution that the morphological classification 
of late-type galaxies is somewhat uncertain. For example, the bar 
classes reported in the LEDA database are different from those in the RC3
in about 20\% (10 out of 47) of the cases. Using the LEDA classification 
scheme, we only find a slightly higher detection rate in barred galaxies 
(Table\,\ref{tab:barstuff}).

The average value of \ione\ in barred galaxies appears slightly higher
than in unbarred galaxies, but given the large standard deviations 
(Col. 7 of Table~\ref{tab:barstuff}) in the respective subsamples,
this is not significant. We have included upper limits for undetected 
galaxies in the averaging, but excluding these values does not change the 
result. For completeness, we point out that the median absolute galaxy 
magnitude (\mb ) of the unbarred subsample is about 1 magnitude fainter 
than for the barred subsample (Col. 8 of Table~\ref{tab:barstuff}) which
potentially causes a bias towards lower molecular gas masses 
in the unbarred sample due to the correlation between \mb\ and \mhtwo .
To summarize, our observations provide little evidence for enhanced
molecular gas masses in the centers of barred late-type spirals.
%
%
\begin{table}
\scriptsize
\caption{Impact of bar class on CO abundance \label{tab:barstuff}}
\begin{tabular}{llcccccc}
\hline
(1) & (2) & (3) & (4) & (5) & (6) & (7) & (8) \\
Database & Class & total & det. & undet. & mean \ione & $\sigma$(\ione) & median \mb \\
& & & & & [$\kkms$] & [$\kkms$]\\
\hline
    & unbarred (SA) & 16 &  7 &  9 & 1.13 & 0.83 & -17.3 \\
RC3 & mixed (SAB)   & 16 & 11 &  5 & 1.56 & 1.50 & -18.2 \\
    & barred (SB)   & 15 & 12 &  3 & 1.80 & 1.80 & -18.4 \\
    & full sample   & 47 & 30 & 17 & 1.49 & 1.42 & -17.8 \\ 

\hline 

     & unbarred	    & 12 &  7 &  5 & 1.30 & 0.90 & -17.3 \\
LEDA & barred       & 35 & 23 & 12 & 1.56 & 1.57 & -18.2 \\
     & full sample  & 47 & 30 & 17 & 1.49 & 1.42 & -17.8 \\ 
\hline
\end{tabular}

Column~7: standard deviation of \ione\ values, including upper limits.
\end{table}

Whether stellar bars are a significant ingredient in the recipe to
form compact stellar nuclei is an even more difficult question. 
Recent observations have not found 
much evidence for enhanced star formation in the very nuclei of 
barred galaxies \citep{ho97}, in apparent contradiction to the
results from molecular gas surveys mentioned above. This notion 
is confirmed by the fact that nuclear clusters are found just 
as often in barred galaxies as in unbarred ones 
(paper I; B\"oker et al. 2003, in preparation).
In addition, the topic is complicated by the fact that bars do not live
forever. In fact, numerical simulations have shown that build-up of a central 
mass concentration (such as a supermassive black hole or a compact nuclear 
cluster) can dissolve stellar bars via dynamical instabilities \citep{nor96}. 
Because the bar destruction happens within just a few dynamical times, 
it is possible that a stellar bar might have been present in the past, 
leading to nuclear star formation until enough mass has been 
collected at the center to render the bar unstable. Without a more
accurate isophotal analysis or even kinematic information on the
central disks of late-type spirals, this question will remain open.
%
\section{Summary} \label{sec:summary}
We have presented \coone\ and \cotwo\ spectra of the central $1-2\kpc$
in 47 spiral galaxies with Hubble-types between Scd and Sm. Of these,
we detect 30 objects in at least one of the lines. Our survey thus 
significantly increases the number of available CO data for very late-type 
disk galaxies. The main results of our analysis can be summarized as follows:
\begin{enumerate}

\item{The average \htwo\ mass within the central kpc is 
$1.4\times10^7\msun$. In principle, this amount is sufficient to 
support a number of modest ($\sim 10^5\msun$ in stars) starburst events. 
As recent high-resolution optical studies of Scd-Sm spirals have shown, 
star formation in these ``pure'' disk galaxies is almost exclusively 
found in the very nucleus, which is the location of a luminous, compact, 
and massive star cluster in at least 75\% of the cases. In this context,
our results are consistent with scenarios that invoke repeated nuclear
starbursts to explain the fact that the SED of many clusters is dominated
by a young (less than a few 100 Myrs) population of stars.}

\item{The latest-type spirals closely follow correlations between
molecular gas content and galaxy luminosity - both at optical and far-infrared
wavelengths - that have been established for more luminous, early-
and intermediate-type spirals and extend them to lower luminosities.
In particular, we find a lower ratio between the (central) 
\mhtwo\ and the (total) \mhi\ for late type spirals
for both prescriptions used to translate CO luminosity 
into molecular gas mass. This could either be due to a lower 
overall CO content (with respect to \mhi) or a more extended distribution
of the molecular gas in late-type galaxies. We favor the second scenario
since the results of \citet{bos02} obtained with a larger
beam indicate that the overall \mhtwo/\mhi\ ratio is the
same in all Hubble types.}

\item{We find a strong separation between our CO detections and 
non-detections in central surface brightness of the stellar disk, $\mu_I^0$.
We detect 93\% of galaxies with $\mu_I^0 < 19\mpas$, but only 8\% of
galaxies with $\mu_I^0 > 19\mpas$. While our observations are not
sensitive enough to establish a direct correlation between the two
quantities, our results fit in well with the detection rate of LSB
galaxies which have even lower surface brightness disks than the galaxies
in our sample. This suggests that the stellar mass distribution of the
galaxy disk is an important indicator for the central accumulation
of molecular gas.} 

\item{While there is little evidence for a direct correlation between the
luminosity of the nuclear star cluster and the \htwo\ mass, we find a 
higher CO detection rate and higher average \htwo\ mass in galaxies with 
the most luminous clusters. Although for young stellar populations 
luminosity is not a reliable indicator of mass, this suggests that the
molecular gas supply is indeed an important parameter for the
formation mechanism of nuclear star clusters.}

\item{Our observations provide only marginal support for an enhanced 
molecular gas mass in the centers of barred galaxies relative to 
unbarred ones.}
\end{enumerate}
%
\acknowledgements
We would like to thank the referee, A. Boselli, for his 
helpful comments.
UL ackowledges support from DGI Grant AYA2002-03338 and the
 Junta de Andaluc\'\i a 
(Spain). This research has made use of the NASA/IPAC Infrared Science Archive
and the NASA/IPAC Extragalactic Database (NED), both of which 
are operated by the Jet Propulsion Laboratory, California 
Institute of Technology, under contract with the National Aeronautics and 
Space Administration. It has also benefited greatly from use of the Lyon-Meudon
Extragalactic Database (LEDA, http://leda.univ-lyon1.fr).


\end{document}